\documentclass[fleqn,usenatbib]{mnras}
\usepackage{newtxtext,newtxmath}
\usepackage[T1]{fontenc}

\DeclareRobustCommand{\VAN}[3]{#2}
\let\VANthebibliography\thebibliography
\def\thebibliography{\DeclareRobustCommand{\VAN}[3]{##3}\VANthebibliography}

\usepackage{graphicx}	    % Including figure files
\usepackage{amsmath}	    % Advanced maths commands
\usepackage{upgreek}        % Upright Greek letters
\usepackage{threeparttable}
\usepackage{tabulary}
\usepackage{caption}
\usepackage{orcidlink}

\newcommand{\DMunits}{\,pc\,cm$^{-3}\,$}
\newcommand{\DMtrendunits}{\,pc\,cm$^{-3}$\,yr$^{-1}$}
\newcommand{\RMunits}{\,rad\,m$^{-2}\,$}
\newcommand{\RMtrendunits}{\,rad\,m$^{-2}$\,yr$^{-1}$}

\newcommand{\driftunits}{\,MHz\,ms$^{-1}\,$}
\def\software#1{\texttt{#1}}

\newcommand{\Swin}{Centre for Astrophysics and Supercomputing, Swinburne University of Technology, P.O. Box 218, Hawthorn, VIC 3122, Australia}

\newcommand{\WIS}{Department of Particle Physics and Astrophysics, Weizmann Institute of Science, Rehovot 7610001, Israel}

\newcommand{\curtin}{International Centre for Radio Astronomy Research, Curtin University, Kent St, Bentley WA 6102, Australia}

\newcommand{\ATNF}{CSIRO Space and Astronomy, Australia Telescope National Facility, PO Box 76, Epping, NSW 1710, Australia}

\newcommand{\GZHU}{Department of Astronomy, School of Physics and Materials Science, Guangzhou University, Guangzhou 510006, China}

\newcommand{\ozgrav}{ARC Centre of Excellence for Gravitational Wave Discovery (OzGrav), Hawthorn, VIC 3122, Australia}

\newcommand{\ZJL}{Research Center for Intelligent Computing Platforms, Zhejiang Laboratory, Hangzhou 311100, China}

\newcommand{\NAOC}{National Astronomical Observatories, Chinese Academy of Sciences, 20A Datun Road, Chaoyang District, Beijing 100101, China}

\newcommand{\XAO}{Xinjiang Astronomical Observatory, Chinese Academy of Sciences, 150 Science 1-Street, Urumqi 830011, China}

\newcommand{\PMO}{Purple Mountain Observatory, Chinese Academy of Sciences, Nanjing 210008, China}

\newcommand{\ASTRON}{ASTRON, Netherlands Institute for Radio Astronomy, Oude Hoogeveensedijk 4, 7991 PD Dwingeloo, The Netherlands}

\newcommand{\JIVE}{Joint institute for VLBI ERIC, Oude Hoogeveensedijk 4, 7991 PD Dwingeloo, The Netherlands}

\title[Repetitions of FRB\,20180301A]{Spectropolarimetric variability in the repeating fast radio burst source FRB\,20180301A}

\author[Kumar et al.]{P.~Kumar\orcidlink{0000-0003-1913-3092},$^{1,2}$\thanks{E-mail: pravir.kumar@weizmann.ac.il}
R.~Luo\orcidlink{0000-0002-4300-121X},$^{3,4}$\thanks{E-mail: rluoastro@gmail.com}
D.~C.~Price\orcidlink{0000-0003-2783-1608},$^{5}$\thanks{E-mail: danny.price@curtin.edu.au}
R.~M.~Shannon\orcidlink{0000-0002-7285-6348},$^{1,6}$
A.~T.~Deller\orcidlink{0000-0001-9434-3837},$^{1}$
S.~Bhandari\orcidlink{0000-0003-3460-506X},$^{7,8}$
\newauthor
Y.~Feng\orcidlink{0000-0002-0475-7479},$^{9}$
C.~Flynn\orcidlink{0000-0003-1110-0712},$^{1}$
J.~C.~Jiang\orcidlink{0000-0002-6465-0091},$^{10}$
P.~A.~Uttarkar\orcidlink{0000-0002-2346-6853},$^{1}$
S.~Q.~Wang\orcidlink{0000-0003-4498-6070},$^{11}$
and
S.~B.~Zhang\orcidlink{0000-0003-2366-219X}$^{12}$
\\
$^{1}$\Swin\\
$^{2}$\WIS\\
$^{3}$\ATNF\\
$^{4}$\GZHU\\
$^{5}$\curtin\\
$^{6}$\ozgrav\\
$^{7}$\ASTRON\\
$^{8}$\JIVE\\
$^{9}$\ZJL\\
$^{10}$\NAOC\\
$^{11}$\XAO\\
$^{12}$\PMO\\
}

\date{Accepted XXX. Received YYY; in original form ZZZ}

\pubyear{2022}

\begin{document}
\label{firstpage}
\pagerange{\pageref{firstpage}--\pageref{lastpage}}
\maketitle

\begin{abstract}
As the sample size of repeating fast radio bursts (FRBs) has grown, an increasing diversity of phenomenology has emerged. Through long-term multi-epoch studies of repeating FRBs, it is possible to assess which phenomena are common to the population and which are unique to individual sources. We present a multi-epoch monitoring campaign of the repeating FRB source 20180301A using the ultra-wideband low (UWL) receiver observations with \textit{Murriyang}, the Parkes 64-m radio telescope. The observations covered a wide frequency band spanning approximately 0.7--4\,GHz, and yielded the detection of 46 bursts. None of the repeat bursts displayed radio emission in the range of 1.8--4 GHz, while the burst emission peaked at 1.1 GHz. We discover evidence for secular trends in the burst dispersion measure, indicating a decline at a rate of $-2.7\pm0.2\,{\rm pc\,cm^{-3}\,yr^{-1}}$. We also found significant variation in the Faraday rotation measure of the bursts across the follow-up period, including evidence of a sign reversal. While a majority of bursts did not exhibit any polarization, those that did show a decrease in the linear polarization fraction as a function of frequency, consistent with spectral depolarization due to scattering, as observed in other repeating FRB sources. Surprisingly, no significant variation in the polarization position angles was found, which is in contrast with earlier measurements reported for the FRB source. We measure the burst rate and sub-pulse drift rate variation and compare them with the previous results. These novel observations, along with the extreme polarization properties observed in other repeating FRBs, suggest that a sub-sample of FRB progenitors possess highly dynamic magneto-ionic environments.
\end{abstract}

\begin{keywords}
methods: data analysis -- methods: observational -- fast radio bursts.
\end{keywords}

%%%%%%%%%%%%%%%%%%%%%%%%%%%%%%%%%%%%%%%%%%%%%%%%%%

\section{Introduction}\label{sec:intro}
Polarization studies of radio signals are vital in probing the immediate environments around the progenitors of astronomical transients. Faraday rotation resulting from the propagation of radio signals through magnetic environments offers a powerful means of probing the magnetic strength and topology along the line of sight \citep{Han:2006, Noutsos:2008}. Radio pulsars, which often exhibit significant linearly polarized radiation, have been used to map the magnetic fields in the interstellar medium of our Galaxy \citep{Sobey:2019}. Exotic variants of pulsars, such as magnetars, have been observed to display large Faraday rotation measures (RMs) of up to~$\sim10^5$\RMunits \citep{Eatough:2013, Shannon:2013}. Furthermore, these objects display significant changes in RM over time, indicating magneto-ionic variations in the progenitor environment \citep{Desvignes:2018}. Long-term polarization studies can thus provide valuable insights into the evolution of the magnetic field properties of these objects over time. Over the past decade, a new class of extragalactic objects, fast radio bursts \citep[FRBs;][]{Lorimer:2007}, has emerged as a means to probe such extreme magneto-ionic environments and, in turn, to aid in mapping the magnetogenesis across galaxies \citep{Vazza:2018}.

FRBs are energetic $\upmu$s--ms-duration radio transients that have been found to originate at up to cosmological distances \citep[see][for recent reviews]{Caleb:2021, Petroff:2022}. The inferred isotropic energies of these transients exhibit a wide range spanning several orders of magnitude from $10^{35}$ to $10^{42}$ erg \citep{Ryder:2022}. There has recently been an exponential growth in the detection of FRB sources, which has established some basic understanding of the statistical properties of these transients \citep{CHIME:2021}. The identifications of their host galaxies (HGs) have also improved our understanding of FRBs, with over two dozen FRB sources localised to HGs, providing the first examinations of the progenitor environments in which these FRBs are produced \citep{Mannings:2021, Bhandari:2022}. The current published FRB source population is dominated by the first discovery catalogue published by the Canadian Hydrogen Intensity Mapping Experiment (CHIME/FRB) telescope \citep{CHIME:2021}. The discovery of FRB-like radio bursts from the Galactic magnetar SGR 1935+2154 \citep{Bochenek:2020, CHIME:2020_magnetar} strengthened the case that at least a portion of FRBs come from magnetars. On the other hand, the discovery of FRB\,20200120E originating from a globular cluster in the nearby galaxy M81 complicates the magnetar origins and suggest that multiple  physical origins for the overall population is possible \citep{Bhardwaj:2021, Kirsten:2022}. 

A critical distinction among the known FRBs is whether or not they exhibit repeat bursts. Only a small fraction ($\sim$\,3 per cent) of them have been found to emit repeat bursts to date \citep{CHIME:2023_repeaters}. The CHIME/FRB catalogue revealed statistical differences in the burst morphology among the two classes of FRBs; the bursts from repeating sources were found with wider pulse widths and narrower spectral extents compared to apparent non-repeating ones \citep{Pleunis:2021}. Furthermore, two well-studied repeating FRB sources exhibit periodic modulation in their burst activity, with a period of 16.33\,d in FRB\,20180916B \citep{CHIME:2020_periodicity, Pleunis:2021_LOFAR} and $\sim$160\,d in FRB\,20121102A \citep{Rajwade:2020, Cruces:2021}. The recurrent nature of repeating FRBs allows for long-term monitoring, which can be utilised to characterise burst properties and their evolutionary trends both in time and across observing frequencies \citep{Xu:2022}. The longstanding question remains whether the repeating and apparently non-repeating FRBs are produced in multiple formation channels or by physically distinct progenitors. The observed distinctions may well be a selection bias due to the beamed emission of FRBs \citep{Connor:2020}. Two of the prolific repeating sources with remarkable similarity in their burst properties, FRB\,20121102A and 20190520B, have been found to be coincident with compact radio continuum sources, known as persistent radio sources (PRS), indicating an existence of a potential sub-group of progenitor environments \citep{Law:2022, Niu:2022}.

Polarization features of the incident burst radiation are one vital tool to characterise the progenitor local environment. While the polarization properties have been measured for a small proportion of the FRB sources, not all of them exhibit polarized radiation. Many are linearly polarized with polarization fractions as high as 100 per cent \citep{Gajjar:2018}. On the other hand, few burst signals are significantly circularly polarized \citep{Cho:2020, Day:2020}. The hallmark properties of repeating FRBs was thought to be the presence of 100 per cent linear polarization (hence the absence of circular polarization) with constant linear polarization position angle (PA) across the burst envelopes \citep{Michilli:2018, Nimmo:2021}. However, recent observations of repeating FRBs have unveiled incredible diversity within the population \citep{Luo:2020}. The discovery of circular polarization in repeating FRBs indicates there may be more in common between one-off and repeating FRB sources \citep{Hilmarsson:2021_CP, Kumar:2022, Xu:2022}. Long-term observations have revealed significant RM evolution in several repeating FRBs, which provides strong evidence for dynamic magneto-ionic environments \citep{Thomas:2022}. This interpretation is further supported by multi-band observations that exhibit a frequency-dependent degree of polarization, consistent with the expected depolarization via multi-path propagation \citep{Feng:2022}.

The 21-cm multibeam (MB) receiver at the Parkes 64-m ``Murriyang'' radio telescope was paramount to early FRB discoveries \citep{Staveley-Smith:1996, Lorimer:2007, Thornton:2013}. The Survey for Pulsars and Extragalactic Radio Bursts (SUPERB) project led these discoveries with the first implementation of a real-time pipeline \citep{Keane:2018}. One such source FRB\,20180301A, was discovered with the Parkes MB during the \emph{Breakthrough Listen} (BL) observations of the Galactic plane \citep{Price:2018, Price:2019}. The burst was detected with a signal-to-noise ratio (S/N) of 16 using the SUPERB FRB pipeline, and was found to have a dispersion measure (DM) of $522\pm5$\DMunits. The burst initially appeared to be confoundingly peculiar and its legitimacy as a bona-fide detection was questioned. The burst spectral bandwidth was measured to be $\sim$138\,MHz confined to a narrow band similar to bursts from other now known repeating FRBs. The burst spectra exhibited circular polarization with an unusually high Faraday RM of $-3163\pm20$\RMunits. The apparent high RM suggested an extreme magnetic environment around the burst progenitor similar to FRB\,20121102A \citep{Michilli:2018}. Soon after the discovery announcement, the FRB source was monitored using telescopes across wavelengths for any repeat bursts. These include follow-up observations at radio wavelengths with the Parkes \citep{Price:2019}. Immediate follow-up campaigns at optical \citep{Xin:2018, Price:2019} and X-ray \citep{Anumarlapudi:2018, Savchenko:2018} wavelengths also resulted in non-detection of a repeat burst or any counterpart emission at the position of the FRB source.

With its unprecedented high sensitivity (a factor of $\sim$80 greater than Parkes in the same band), the Five-hundred-meter Aperture Spherical radio Telescope (FAST) can uniquely search for the lowest energy repetitions from an FRB source. Observations with the FAST 19-beam receiver centred at 1.25\,GHz during 2019 July--October yielded detections of 15 repeat bursts from the FRB\,20180301A source in a total observing time of 12\,h \citep{Luo:2020}. The FAST-detected bursts displayed a variety of dynamic pulse structures, including the well-established downward drifting in frequency among the repeating bursts \citep{Hessels:2019}. A diversity of polarimetric properties was also observed in these bursts, including the first evidence of significant variation in the linear polarization PA across the pulse profiles of repeating FRBs. The high degree of linear polarization and no detection of a circular component were consistent with the most repeating FRBs. Burst-to-burst RM variation with a standard deviation of 14\RMunits was also reported in the FAST-detected bursts around an average RM of 542\RMunits.

\begin{figure*}
\centering
\includegraphics[width=\textwidth]{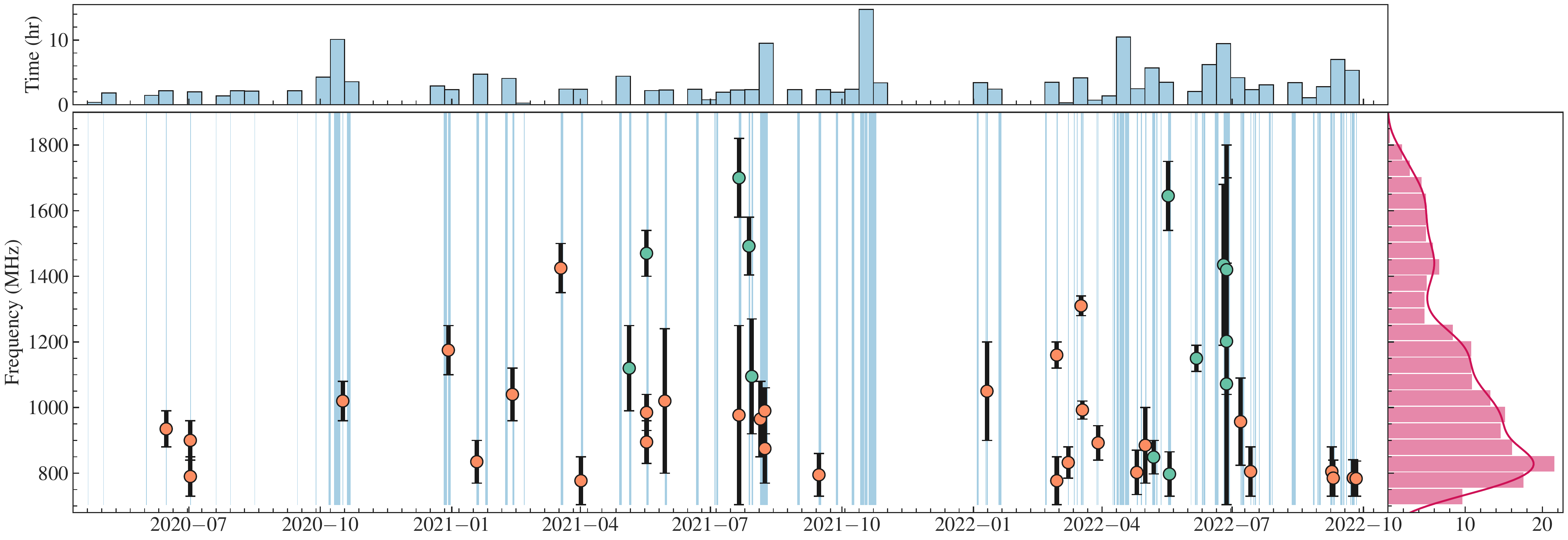}
\caption{Timeline of the Parkes/UWL follow-up observations of FRB\,20180301A source, along with the  detected repeat bursts. {\em Bottom Panel}: Follow-up duration is displayed as shaded regions and magnified for visual representation in the left sub-plot. The burst markers are over-plotted at the zoom level, with green and orange markers indicating bursts with and without a detection of polarization, respectively. The error bars represent the spectral envelope of the individual bursts. The right sub-plot shows the histogram of spectral coverage of the repeat bursts, assuming uniform intensity. {\em Top Panel}: Follow-up duration is displayed as a bar plot binned in intervals of 10\,d.}
\label{fig:timeline}
\end{figure*}

The detection of repetitions enabled the FRB\,20180301A source to be localised using the Karl G. Jansky Very Large Array (VLA), and subsequently associated with the star-forming galaxy PSO\,J093.2268+04.6703 at a redshift of $z=0.33$ \citep{Bhandari:2022}. No persistent radio emission coinciding with the FRB position was found in the VLA observations. A more recent multi-wavelength campaign in 2021 March detected five repeat bursts with FAST but no simultaneous emission in the X-ray band \citep{Laha:2022}. The prolific repeating nature of FRB\,20180301A source makes it an ideal candidate for long-term monitoring to study the evolution of the burst properties. Wideband instrument observations are necessary to determine the chromaticity of the FRB emission. The band-limited nature of repeating FRBs presents challenges in obtaining a robust estimation of the burst rate, as detection efficiency can be considerably affected by the characteristics of the instrument and search pipeline. The peak emission frequency for repeating sources also appears to be variable \citep{Kumar:2021}. This results in a severe completeness limit across the observing band in instruments with narrow bandwidth \citep{Agarwal:2020}. On the other hand, conducting simultaneous observations with multiple telescopes to cover different parts of the spectrum have its logistical challenges \citep{Law:2017}. Thus, the ultra-wideband low (UWL) receiver at Parkes is the ideal instrument for FRB monitoring campaigns.

In this paper, we report the results from a multi-year monitoring campaign of the repeating FRB source 20180301A using the Parkes radio telescope. The observations probed some of the high-activity periods of the FRB source across a broad ($\approx 3$\,GHz) spectral band. The follow-up observations and search methods used to find repeat bursts are described in Section~\ref{sec:observations}. Burst spectro-temporal and polarization properties are presented and modelled in Section~\ref{sec:repeats}. Finally, in Section~\ref{sec:discussion}, we discuss the measured burst properties, their long-term variations and implications on the emission mechanism. %of the FRB source.

\section{Observations and data processing}\label{sec:observations}
We started monitoring the sky position of FRB\,20180301A in 2020 April with the UWL receiver at the Parkes radio telescope. The low-frequency radio instrument provides a continuous frequency coverage from $704$ to $4032$~MHz. We channelised the bandwidth of UWL data into 13312 ($250$-kHz resolution), 3328 ($1$-MHz resolution) and 6656 ($500$-kHz resolution) frequency channels during 2020 April--September, 2020 October--2021 October and 2022 January--September, respectively owing to data storage constraints during the respective periods. We recorded the data at a time resolution of $64$~$\upmu$s with each channel coherently dedispersed to a DM of 517\DMunits \citep{Hobbs:2020}. The data were stored in 8-bit sampled \software{psrfits} search-mode file with full Stokes information being recorded \citep{psrchive}. For the initial observations, we used the source position determined by FAST detections i.e. R.A. = $06^{\rm{h}}12^{\rm{m}}54.96^{\rm{s}}$ and Decl. = $+04\degr38\arcmin43.6\arcsec$ (J2000.0 epoch) \citep{Luo:2020}. Since 2022 January, we have been monitoring the FRB source using the more precise position obtained interferometrically with VLA, which has an offset of 1.5 arcmin from our earlier position \citep{Bhandari:2022}. This 1.5 arcmin offset is well within the primary beam of the telescope (7 arcmin at 4.0 GHz) at the highest frequency observed. Hence the sensitivity of the observations during 2020--2021 are not compromised and no corrections to the burst flux density during these epochs were made.

\begin{table}
\centering
\caption{Example of the tiered sub-band search configuration used for the Parkes/UWL data with a frequency resolution of $500$\,kHz.} 
\label{tab:subband_params}
\begin{threeparttable}
\begin{tabulary}{1\columnwidth}{CCCCC}
\hline  \hline
Sub-bands & Width$_\mathrm{sub}$ & DM & DM trials & Sensitivity\tnote{$\dagger$}\\
  & (MHz) & tolerance &   & (Jy ms)\\
\hline
1  & 3328 & 1.40   & 3091       & 0.15\\
2  & 1664 & 1.16   & 1710--3270 & 0.21\\
4  & 832  & 1.05   & 1240--3121 & 0.30\\
8  & 416  & 1.012  & 1142--3307 & 0.42\\
13 & 256  & 1.005  & 1061--3313 & 0.54\\
26 & 128  & 1.002  & 868--3097  & 0.76\\
52 & 64   & 1.001  & 726--3291  & 1.08\\
\hline
\end{tabulary}
\begin{tablenotes}[flushleft]
    \item[$\dagger$]{The limiting fluence for a pulse width of 1~ms and S/N threshold of 10$\sigma$ assuming that the burst signal occupies the whole sub-band.}
\end{tablenotes}
\end{threeparttable}
\end{table}

\begin{figure*}
\centering
\includegraphics[width=\textwidth]{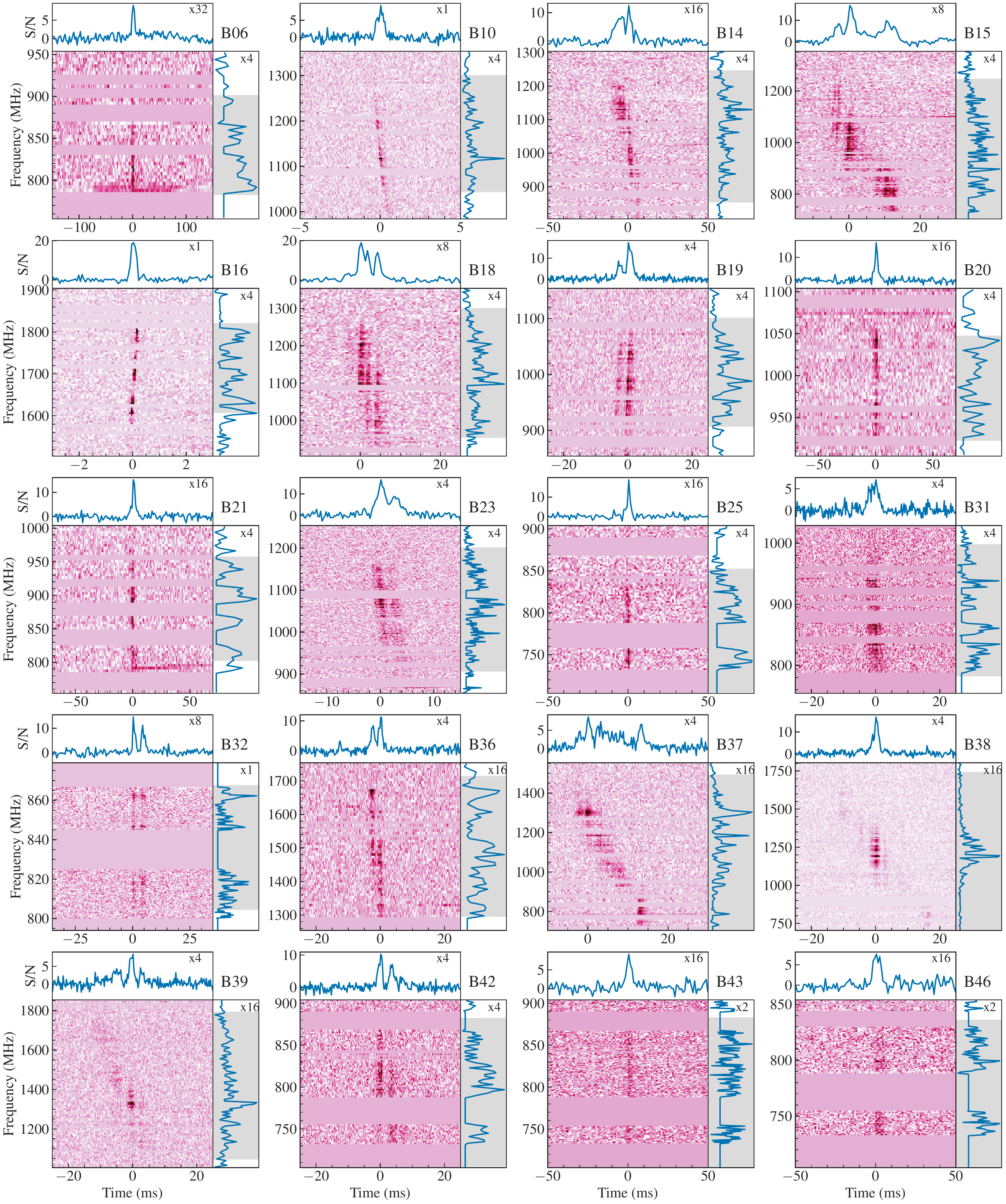}
\caption{
Dynamic spectra of a selection of bright Parkes-detected repeat bursts from the FRB\,20180301A source. Burst arrival times are provided in Table~\ref{tab:burst_properties}. Each burst has been dedispersed to the global source DM of its respective burst group; 517.2 \DMunits for B06--B21 and 514.7 \DMunits for B23--B46. In each panel, the bottom sub-plot displays the dynamic spectrum, and the right sub-plot shows the time-averaged on-pulse spectrum. The downsampling factor for frequency and time resolutions are indicated in the respective sub-plots. The grey band in the right sub-plot of each panel indicates the best-fitting spectral sub-band. The top sub-plot displays the frequency-averaged pulse profile using the best-fitting sub-band. All dynamic spectra are normalized and intensity values are saturated at the fifth percentile to enhance visibility.}
\label{fig:repeaterplots}
\end{figure*}

\begin{table*}
\caption{Burst parameters for FRB\,20180301A repetitions detected with the Parkes/UWL. Except for the detection S/N, all other burst properties are measured for the best-fitting sub-band after dedispersing to the DM $=517.2$\DMunits (for B01--B22) and $=514.7$\DMunits (for B23--B39).} 
\label{tab:burst_properties}
\small
\centering
\begin{threeparttable}
\begin{tabulary}{\textwidth}{LCCCCCCCRCRR}
\hline
Burst & TNS event & TOA\tnote{(a)} & S/N$_{\rm det}$\tnote{(b)} & DM$_\mathrm{S/N}$\tnote{(c)} & DM$_\mathrm{struct}$\tnote{(d)} & $\nu_\mathrm{low}$\tnote{(e)} & $\nu_\mathrm{high}$ & Width\tnote{(f)} & S/N$_{\rm int}$\tnote{(g)} & Fluence\tnote{(h)} & S\tnote{(i)} \\
& FRB & (MJD) & & \DMunits & \DMunits & (MHz) & (MHz) & (ms) &  & (Jy ms) & (Jy) \\
\hline
B01 & 20200615E & 59015.11443196 & 9.7  & 515(1)   & -        & 880(10)  & 990  &  17(3)  & 9.1  & 3.3(4) & 0.5(1)\\
B02 & 20200701I & 59031.99032290 & 10.5 & 518(2)   & -        & 840      & 960  &  15(3)  & 9.6  & -      & -\\
B03 & 20200702E & 59032.04080996 & 9.6  & 515(1)   & -        & 773(17)  & 850  &  2.9(4) & 11.8 & -      & -\\
B04 & 20201016B & 59138.68765679 & 11.7 & 518(1)   & -        & 960      & 1080 &  3.3(4)  & 13.4 & 2.7(3) & 0.8(1)\\
B05 & 20201229E & 59212.57645719 & 9.2  & 518(2)   & -        & 1100     & 1250 &  2.8(4)  & 8.3  & 2.4(3) & 0.8(1)\\
B06 & 20210118D & 59232.48841686 & 13.9 & 518(1)   & 518.4(4) & 770(20)  & 900  &  5.0(4)  & 15.8 & 6.9(4) & 1.5(2)\\
B07 & 20210212G & 59257.40224009 & 9.7  & 519(2)   & -        & 960      & 1120 &  5(1)    & 10.4 & 2.2(3) & 0.6(1)\\
B08 & 20210318A & 59291.28023142 & 9.1  & 519(2)   & -        & 1350     & 1500 &  2.5(3)  & 13.4 & 1.8(1) & 0.7(1)\\
B09 & 20210401C & 59305.37794153 & 8.1  & 518(1)   & -        & 704*     & 850  &  3(2)    & 7.2  & 2.9(4) & 0.9(2)\\
B10 & 20210505H & 59339.14681265 & 13.9 & 517.7(1) & 517.7(1) & 990      & 1250 &  0.20(1) & 28.7 & 0.99(4) & 5.2(3)\\
B11 & 20210517A & 59351.26398778 & 9.1  & 517(2)   & -        & 1400     & 1540 &  10(2)   & 7.3  & 2.1(3) & 0.5(1)\\
B12 & 20210517B & 59351.28957987 & 8.4  & 518(2)   & -        & 930      & 1040 &  3.3(5)  & 8.6  & 2.2(3) & 0.6(1)\\
B13 & 20210517C & 59351.30674364 & 11.6 & 519(2)   & -        & 830      & 960  &  9(2)    & 8.1  & 4(1) & 0.8(2)\\
B14 & 20210530F & 59364.07408752 & 38.1 & 523.4(4) & 518.0(4) & 800      & 1240 &  12(1)   & 24.3 & 6.4(3) & 0.8(1)\\
B15 & 20210720F & 59415.94033215 & 43.5 & 521.3(4) & 516.5(2) & 704*     & 1250 &  18(1)  & 33.0 & 10.5(4) & 1.2(1)\\
B16 & 20210720G & 59415.97642127 & 23.7 & 516.6(3) & 516.6(1) & 1580     & 1820 &  0.09(1) & 49.8 & 0.88(2) & 8.8(2)\\
B17 & 20210727A & 59422.92217547 & 11.9 & 519(2)   & -        & 1404     & 1580 &  2.4(3)  & 13.2 & 1.5(1) & 0.7(1)\\
B18 & 20210729D & 59424.90708801 & 24.9 & 519.2(8) & 517.5(1) & 920      & 1270 &  6.2(3)  & 39.8 & 7.2(2) & 1.3(1)\\
B19 & 20210804A & 59430.90001027 & 36.1 & 516.4(4) & 516.1(2) & 850      & 1080 &  2.2(1)  & 37.6 & 5.6(1) & 2.6(1)\\
B20 & 20210808C & 59434.01279889 & 11.9 & 518.5(8) & 516.6(3) & 920      & 1060 &  2.5(2)  & 21.8 & 4.2(2) & 1.6(1)\\
B21 & 20210808D & 59434.07006206 & 12.6 & 517.8(8) & 518.2(4) & 770(20)  & 980  &  2.3(2)  & 20.3 & 4.7(3) & 1.6(1)\\
B22 & 20210914D & 59471.90486661 & 8.1  & 519 (7)  & -        & 730      & 860  &  20(4)   & 7.2  & 9(1)   & 0.6(1)\\
B23 & 20220110A & 59589.48930034 & 41.4 & 516.6(4) & 515.6(2) & 900      & 1200 &  3.9(1)  & 45.3 & 7.3(2) & 1.9(1)\\
B24 & 20220228G & 59638.31130584 & 12.8 & 516(1)   & -        & 1120     & 1200 &  1.6(2)  & 11.7 & 3.4(3) & 1.8(3)\\
B25 & 20220228H & 59638.34137886 & 18.9 & 515.2(4) & 515.1(2) & 704*     & 850  &  4.0(3)  & 15.7 & 6.4(4) & 1.6(2)\\
B26 & 20220308F & 59646.36674425 & 10.4 & 515(1)   & -        & 785*     & 880(10)  &  6.0(6)  & 14.0 & 5.2(4) & 1.3(2)\\
B27 & 20220317F & 59655.35058910 & 7.5  & 519(1)   & -        & 1280     & 1340 &  0.4(2) & 7.4  & -      & -\\
B28 & 20220318D & 59656.27841775 & 8.0  & 517(1)   & -        & 965      & 1020 &  0.9(2) & 7.1  & 1.2(2) & 1.2(2)\\
B29 & 20220329D & 59667.25739036 & 12.7 & 518(2)   & -        & 840      & 945  &  6.5(7) & 13.1 & 7.1(6) & 1.2(2)\\
B30 & 20220425B & 59694.33193415 & 9.3  & 516(1)   & -        & 735      & 870  &  2.3(4) & 8.4  & 2.4(3) & 1.2(2)\\
B31 & 20220501D & 59700.28893142 & 33.1 & 515.4(6) & 515.3(3) & 770(20)  & 1000 &  3.3(2)  & 19.9 & 8.8(5) & 2.8(3)\\
B32 & 20220507B & 59706.10269798 & 19.4 & 515(1)   & 514.8(1) & 798*     & 900(30)  &  7.0(6)  & 20.4 & 8.4(5) & 3.1(3)\\
B33 & 20220517A & 59716.31087602 & 10.8 & 518(1)   & -        & 1540     & 1750 &  1.4(2)  & 10.0 & 0.7(1) & 0.5(1)\\
B34 & 20220518A & 59717.20933229 & 21.2 & 515(1)   & 515.4(2) & 730*     & 865  &  11(1)   & 13.8 & 6.1(6) & 1.2(2)\\
B35 & 20220606A & 59736.06144291 & 9.7  & 515(1)   & -        & 1110     & 1190 &  1.4(3)  & 7.9  & 1.8(2) & 1.1(2)\\
B36 & 20220625A & 59755.10979008 & 25.1 & 517.5(7) & 514.1(1) & 1190(110)& 1680 &  4.1(3)  & 24.9 & 2.4(1) & 0.9(1)\\
B37 & 20220627A & 59757.08891623 & 45.8 & 518.0(4) & 518.6(3) & 704*     & 1440 &  16(1)   & 27.2 & 7.7(3) & 0.8(1)\\
B37A & -        & -              & -    & 518.7(4) & 519.0(2) & 920      & 1360 &  1.99(3)  & 92.7 & 8.8(1) & 5.1(2)\\
B37B & -        & -              & -    & 514.8(3) & 514.7(2) & 704*     & 920  &  1.99(3)  & 92.7 & 8.8(1) & 5.1(2)\\
B38 & 20220627B & 59757.12153219 & 98.9 & 514.5(7) & 514.8(2) & 704*     & 1700 &  2.1(1)  & 33.3 & 3.7(1) & 1.7(1)\\
B38A & -        & -              & -    & 515.0(4) & 514.9(2) & 1004 & 1404 &  1.99(3)  & 92.7 & 8.8(1) & 5.1(2)\\
B39 & 20220627C & 59757.18310120 & 26.2 & 520(1)   & 514.3(2) & 1040     & 1800     & 10(1)  & 20.2 & 3.1(2) & 0.6(1)\\
B40 & 20220706A & 59766.96164601 & 8.6  & 516(1)   & -        & 824      & 1090(10) & 2.1(2) & 11.9 & 2.6(2) & 1.2(2)\\
B41 & 20220713A & 59773.97654888 & 23.2 & 516(1)   & 515.1(5) & 770(15)  & 880(10)  & 2.3(3) &  11.4 & 5.8(5) & 2.9(4)\\
B42 & 20220908A & 59830.79113887 & 22.1 & 516(1)   & 514.6(2) & 730*     & 880(10)  & 7.8(6) & 21.2 & 8.3(4) & 2.1(1)\\
B43 & 20220908B & 59830.80702725 & 10.5 & 515(1)   & -        & 730*     & 880(10)  & 3.7(3) & 15.6 & 4.1(3) & 1.7(2)\\
B44 & 20220909A & 59831.80415599 & 12.1 & 515(1)   & -        & 730*     & 840(5)   & 3.4(4) & 11.8 & 3.9(4) & 1.5(2)\\
B45 & 20220923A & 59845.82918222 & 11.0 & 514(1)   & -        & 730*     & 841(2)   & 1.8(2) & 12.1 & 2.4(2) & 1.4(2)\\
B46 & 20220925A & 59847.82634130 & 14.6 & 514(1)   & 514.3(5) & 730*     & 837(7)   & 5.2(5) & 14.4 & 6.0(5) & 1.2(2)\\
\hline
\end{tabulary}
\begin{tablenotes}[flushleft]
    \item[(a)] {Time of arrival in Barycentric Dynamical Time (TDB) at $\nu=\infty$}
    \item[(b)] {Reported S/N by the \software{heimdall} search pipeline (for the best sub-band detection).}
    \item[(c)] {DM obtained by maximizing S/N using \software{pdmp}. DM uncertainties are 1$\sigma$ conservative estimates.}
    \item[(d)] {DM obtained by maximising the coherent power across the bandwidth with \software{DM\_phase}.}
    \item[(e)] {For highlighted bursts (*), the spectral envelope extends beyond the telescope bandwidth, and a lower limit could not be obtained.}
    \item[(f)] {Burst width (FWHM), obtained by fitting the pulse profile with a Gaussian function.} 
    \item[(g)] {Pulse profile is convolved with normalized Gaussian filters (boxcar filter for bursts with multiple sub-components) to calculate integrated S/N.}
    \item[(h)] {Bursts B02, B03 and B27 lack calibrated flux measurements due to corrupt noise diode observations.}
    \item[(i)] {Peak flux density. Uncertainties for fluence and peak flux density are calculated from off-pulse noise.}
\end{tablenotes}
\end{threeparttable}
\end{table*}

\begin{figure*}
\centering
\includegraphics[width=\textwidth]{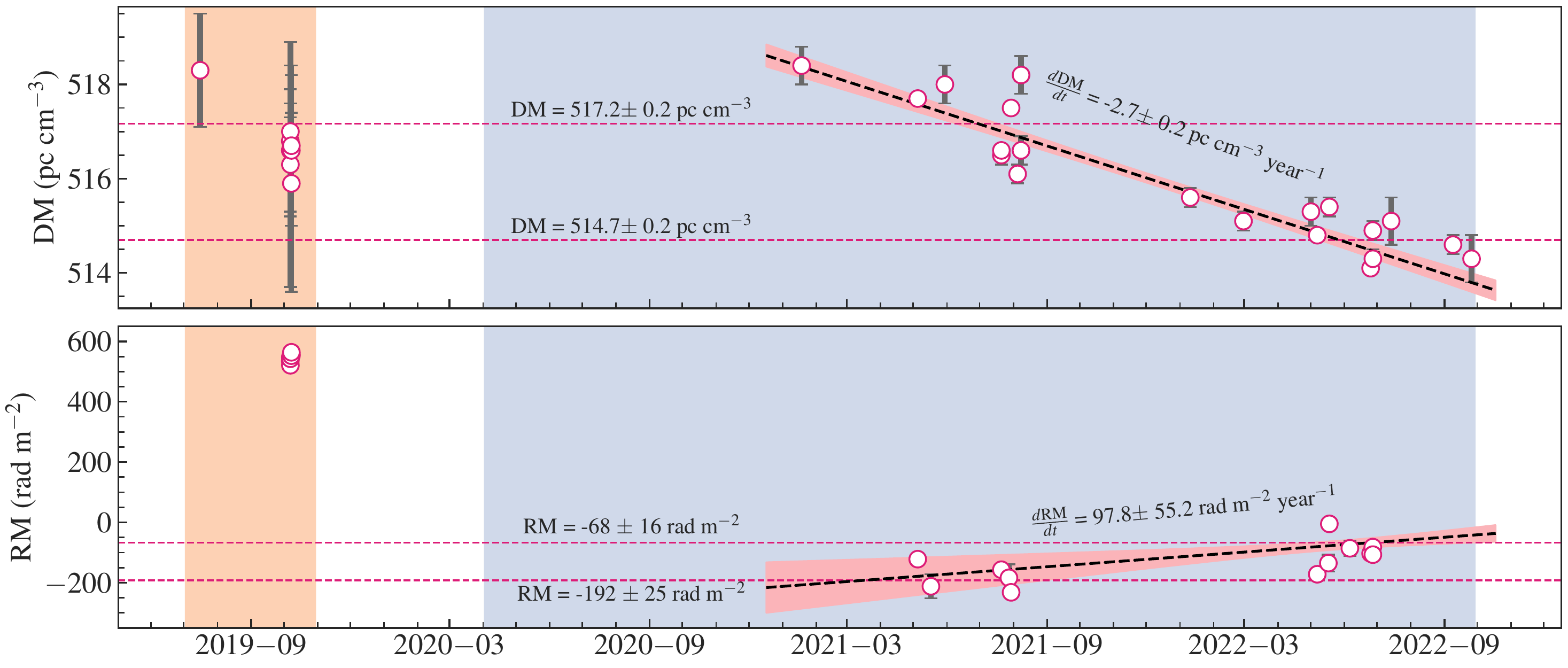} 
\caption{DM and the conventional Faraday RM variations of the FRB\,20180301A bursts. The follow-up span with Parkes/UWL and FAST \citep{Luo:2020} is displayed as a shaded region in blue and orange, respectively. The dashed red line indicates the weighted mean for the two sub-groups, G1 and G2 (see Section~\ref{sec:dm}). The dashed black line represents the best-fitting line, obtained by fitting a linear model to the DM$_\mathrm{struct}$ (decreasing trend) and RM$_\mathrm{nest}$ (increasing trend) for the UWL bursts (excluding B37). The shaded background is the 1$\sigma$ uncertainty.}
\label{fig:dm_rm_evolution}
\end{figure*}

We monitored the FRB source position for a total of 193\,h distributed in 120 separate observing sessions across the time span of $2.5$\,yr. To ensure accurate polarization calibration, we conducted brief 1--2-min observations of a pulsed linearly polarized noise diode at the beginning of each session. We then corrected for polarimetric leakage using an instrumental response model for UWL provided by the observatory. The response model was derived from extensive observations of the millisecond pulsar PSR~J0437$-$4715 \citep{Johnston:1993} covering a wide range of parallactic angles \citep{vanStraten:2004}. For flux calibration, we relied on observations of the radio quasar PKS~0407$-$658, which were also provided by the observatory, to relate the amplitude of the noise diode to a physical scale. We performed single-pulse searches in the UWL data using a tiered sub-band strategy to detect possible narrow-band astrophysical signals \citep{Kumar:2021}. The recorded data was divided into equal sub-bands with sizes ranging from 128 to 3328\,MHz. For data with higher frequency resolution, we also probed smaller scale sub-bands with a band size of 64\,MHz. To ensure sensitivity to bursts that overlapped sub-band boundaries, we formed additional sub-bands by joining the bottom and top half portions of the adjacent sub-bands for each search tier.

We employed a GPU-based single-pulse search software \software{heimdall\footnote{\url{https://sourceforge.net/projects/heimdall-astro/}}} to search the sub-banded data with a fixed DM range of 100--1100 \DMunits \citep{Barsdell:2012PhDT}. The trial DM step sizes used in the analysis were internally calculated by \software{heimdall} using an algorithm that accounts for the pulse broadening from the previous DM step \citep{Levin:2012PhDT}. This effect is encapsulated in the DM tolerance parameter, which determines the average S/N loss for the next trial DM \citep{Rane:2016}. Table \ref{tab:subband_params} lists the DM tolerance values used for each sub-band search configuration and the resulting number of DM trials for $500$-kHz resolution data. To process the sub-banded data using \software{heimdall}, we created a total intensity bandpass-corrected \software{sigproc} filterbank \citep{Lorimer:2011} file from the telescope data using \software{digifil} \citep{dspsr}. We flagged channels affected by radio frequency interference (RFI) with a modified Z-score greater than $5.0$. The standard score for frequency channels is computed using the median absolute deviation of their statistical moments, up to kurtosis.  We applied the following criteria to filter the clustered candidates obtained from \software{heimdall}: S/N > 7.5, 0.064 ms < pulse width < 32.768 ms and cluster members > 15. 

We utilized \software{FETCH}, a pre-trained convolutional neural network model, to perform the classiﬁcation of the real astrophysical signal from RFI among the signal candidates \citep{Agarwal:2020}. We employed their best-trained model, \software{A}, with a probability threshold of 0.5 for classification. \software{FETCH} identified a total of 30761 candidates as real signals, from which we manually identified 306 as genuine, consisting of multiple sub-band detections of the same burst. In addition, we visually inspected all 69906 candidates with DM within $\pm20$\DMunits of the FRB DM and a S/N above $8$ to confirm the \software{FETCH} classifications. Our analysis of the Parkes observations yielded a total of 46 signal candidates identified as repeat bursts. Four repeat bursts with S/N < 10 were only detected through manual inspection and not by \software{FETCH}. Notably, we did not detect any FRB candidates with DM inconsistent with that of the FRB 20180301A source.

\section{Results}\label{sec:repeats}
The Parkes-detected repeat bursts have considerable morphological diversity in their dynamic spectra. No bursts show emission above 1.8\,GHz, with most of them being in the lower part of the band. Hereafter, we refer to the repeat bursts chronologically as B01--B46. In Fig.~\ref{fig:timeline}, we show the timeline of Parkes observations, including the detected repeat bursts and their spectral coverage. Figure~\ref{fig:repeaterplots} displays the dynamic spectra of repeat bursts with high S/N (>15). We extract the data around the burst signal with complete Stokes information in a \software{psrfits}-format file using the \software{dspsr} package \citep{dspsr} to measure the burst properties. To minimize the impact of RFI and other noise sources, we mask frequency channels whose statistical moments deviated significantly from the median values per channel across the entire band. For this, we utilize the list of channels flagged during the burst search procedure calculated for the whole data file of duration $\sim$30\,s. Additionally, we manually identify and mask residual RFI-affected channels in some of the bursts. We calibrate the extracted data using the \software{pac} tool from \software{psrchive} \citep{psrchive}. We use noise diode observations to correct for the differential gain and phase of the telescope receptor polarization. In the UWL calibrator data provided by the observatory, frequency channels known to be highly affected by RFI, as well as those affected by aliasing at the edges of the $128$~MHz sub-bands were excised by default \citep{Hobbs:2020}. To address the gaps in the data resulting from this removal, we used the tool \software{smint} from \software{psrchive} to interpolate and smooth the flux and polarization calibrator solutions before applying them to the burst data.

\subsection{Dispersion measure}\label{sec:dm}
The dynamic spectrum of FRB signals commonly exhibits complex features intrinsic to the pulse mimicking dispersive effects distinct from the cold-plasma dispersion relation $\nu^{-2}$ \citep{Hessels:2019}. These features are usually interpreted in terms of the frequency-drifting of the burst components and are commonly referred to as the ``sad-trombone'' effect \citep{Hessels:2019}. Additionally, some bursts exhibit bifurcating and non-alignable structures in parts of the dynamic spectral region resulting in distinct DM estimates for the individual sub-components \citep{Hessels:2019, Marthi:2020, Platts:2021, Kumar:2022}. These could be due to the unresolved sub-structures in the burst owing to the limited time resolution \citep{Platts:2021}. Nevertheless, one thing is clear: an estimate of DM maximizing the frequency-averaged S/N often blurs the dynamic burst structure and is thus inappropriate for burst signals with complex morphology and sub-components. Hence, estimating the DM by maximizing the signal structure is the accepted practice for accurately quantifying the cold plasma dispersive effects in bursts with significant structure. However, for bursts with low S/N, it is often not possible to determine a reliable structure maximising DM.

We therefore use both the S/N and structure maximization approaches to determine the best DM for each repeat burst. We identify the burst spectral band by eye based on the signal brightness. We also take a broader sub-band (band increment up to 20\,MHz) to confirm the spectral coverage and assess whether our measurements are robust to choice of sub-band. We conduct the DM analysis at different resolutions in time and frequency and report the outcome with the highest S/N. We perform a boxcar matched-filtering of the pulse profiles for S/N maximization using the \software{PSRCHIVE} tool \software{pdmp}. We find a wide spread in the obtained DM measurements, ranging from 515 to 524\DMunits in the UWL bursts, which have been observed before in repeating bursts with a complex dynamic spectrum. Some of the bursts (e.g., B14, B15, B18, B36--B39) have complex and distinct sub-components (see Fig.~\ref{fig:repeaterplots}). To align the burst structure accurately in time and account for these sub-components, we use the \software{DM\_phase} package \citep{Seymour:2019}. Unlike traditional methods that rely on maximizing S/N, this method utilizes the coherent power across the dynamic spectrum to optimize the signal structure. We adopted a polynomial fitting approach, as described in \software{DM\_phase} \citep{Seymour:2019} to estimate the DM uncertainties for both methods. Specifically, we fitted the obtained S/N-DM curve \citep{Cordes:2003} with a polynomial of higher degree (10), and extracted the DM value at the peak of the fit. The associated uncertainty was calculated from the residuals of the polynomial fitting \citep{Platts:2021}. The DM measurements are reported in Table \ref{tab:burst_properties}. For bursts with a low S/N ($\lesssim 14$), the structure-maximization approach does not result in good fit of the S/N-DM curve, so we do not provide measurements for them. From now on, we only consider the DM$_\mathrm{struct}$ as the measurement of the source DM.

Our analysis of the UWL repeat bursts reveals compelling evidence of a secular, long-term decrease in the structure-maximized DM values, particularly in the bursts detected during 2021--2022. We found no significant evidence for DM variations with respect to the burst emission frequency. To quantify the temporal trend, we applied a linear regression model to the DM evolution, which yielded a gradient of
\begin{equation}
    \frac{d\rm DM}{dt} = -2.7\pm0.2\,{\rm pc\,cm^{-3}\,yr^{-1}},
\end{equation}
indicating a gradual decrease over time. However, it is important to note that the linear regression resulted in a reduced $\chi^2$ statistic of $\chi^2_{\nu} \approx 7.5$, suggesting that the model does not provide an optimal fit to the data. The presence of outlier DM measurements, leading to significant residuals, raises the possibility of unresolved components in the burst structures. Nonetheless, we focus on interpreting the long-term secular trend observed in the data.
The measured DM$_\mathrm{struct}$ of the Parkes bursts with their uncertainties, including the linear fit for the decreasing trend, are shown in the top panel of Fig.~\ref{fig:dm_rm_evolution}. For reference, we also show the structure-optimized DM measurements of the FAST-detected bursts from 2019 \citep{Luo:2020}. The DM of bursts from the FRB\,20180301A has significantly changed during the three years of Parkes monitoring. We find that a single measurement of DM is unsuitable for obtaining burst properties for the entire Parkes/UWL sample. The bursts are visibly under/over-corrected for dispersion using any such average measurement. For convenience, we divide the UWL burst sample into two groups, G1 (bursts detected in 2020--2021) and G2 (2022 onwards). The weighted mean of DM$_\mathrm{struct}$ measurements for repeat bursts during the respective periods can be used to define a global group DM for this FRB source. We thus obtain the group DM to be $517.2 \pm 0.2$ \DMunits for bursts in G1 and $514.7 \pm 0.2$ \DMunits for G2 (excluding B35, see below). The inter-group slope indicates that the average FRB source DM for bursts in G2 has decreased by 0.5 per cent with more than $12\sigma$ significance compared to G1 bursts.

\begin{figure}
\centering
\includegraphics[width=\columnwidth]{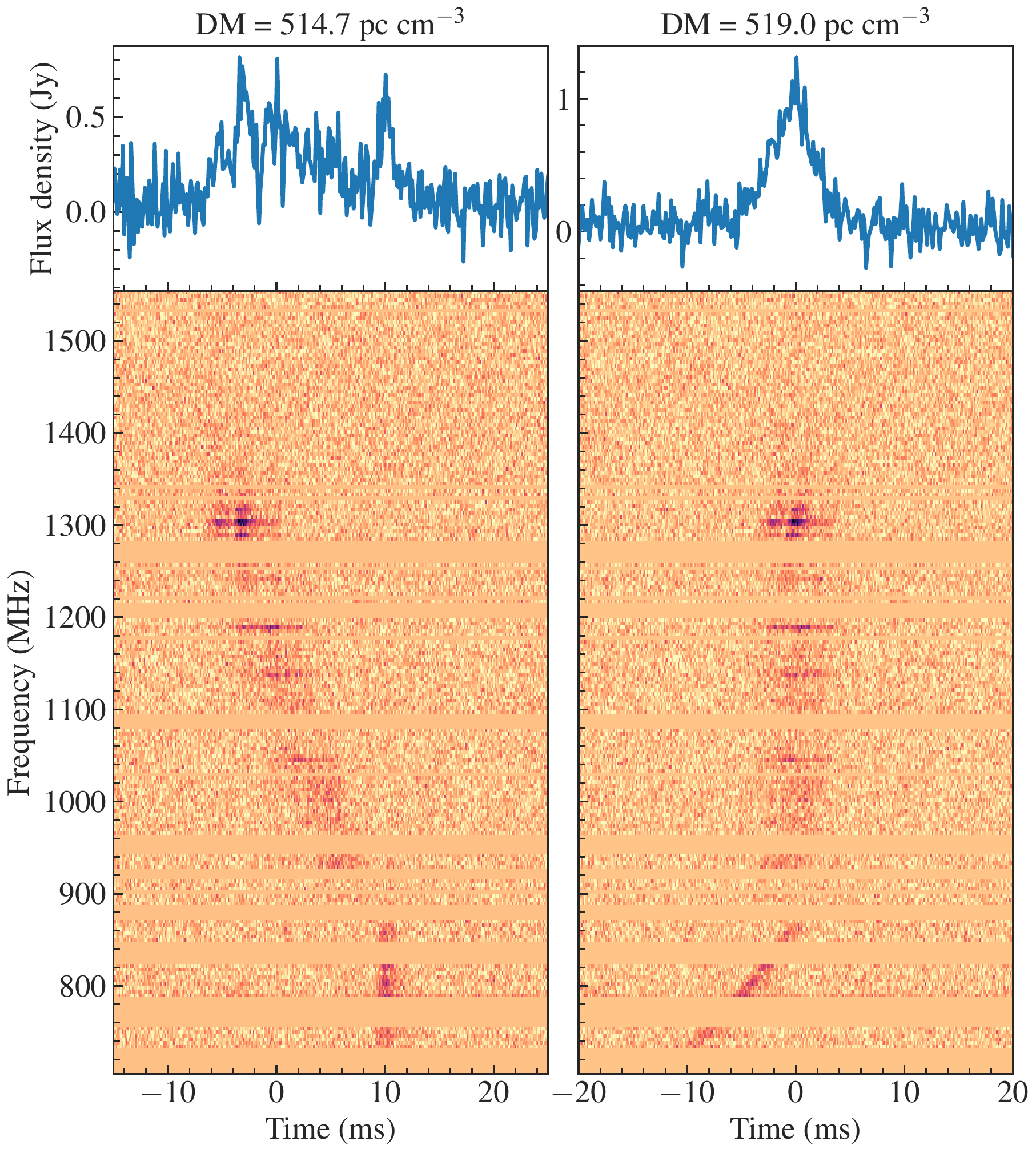} 
\caption{Intensity dynamic spectra of the UWL repeat burst B37, dedispersed to the DM of 514.7 and 519\DMunits. The two DM scenarios correspond to the alignment of the low and high-frequency components of the burst signal in time, respectively. The normalized spectra are plotted with 2\,MHz spectral resolution and 1\,ms temporal resolution, with values capped at the 95th percentile. The frequency-averaged pulse profiles are displayed in the top sub-plots of each panel.}
\label{fig:dm_comp}
\end{figure}

While the DM analysis presented above is straightforward, there are several caveats. To begin, we only used the DM measurements of the Parkes/UWL detections to maintain consistency and limit any potential instrumental biases. \cite{Luo:2020} calculated and maximized the dedispersed pulse profile's S/N-weighted local contrast to achieve the best alignment. On the other hand, we use the more robust method of optimising the burst power in the Fourier domain for the Parkes bursts \citep{Seymour:2019}. Unlike \cite{Luo:2020}, who used the spread of the contrast-DM curve to estimate DM uncertainty, we use the best-fit residuals of the S/N-DM curve to estimate uncertainty \citep{Platts:2021}. As a result, the associated uncertainties are distinct and cannot be compared. Second, the FRB source may have different regimes or phases of DM variations. This could be one of the reasons why FAST DM measurements do not fit in with the overall secular trend. Finally, there has yet to be a consensus established to estimate the uncertainty with the maximizing structure approach. We use the conventional method here to maintain consistency with results from other repeating FRBs. Even if the DM errors are underestimated, a trend in the best-fitting DM values can still be seen. Furthermore, it is clear that using the earlier DM estimates results in a misaligned burst structure for the latter bursts.

\begin{figure}
\centering
\includegraphics[width=\columnwidth]{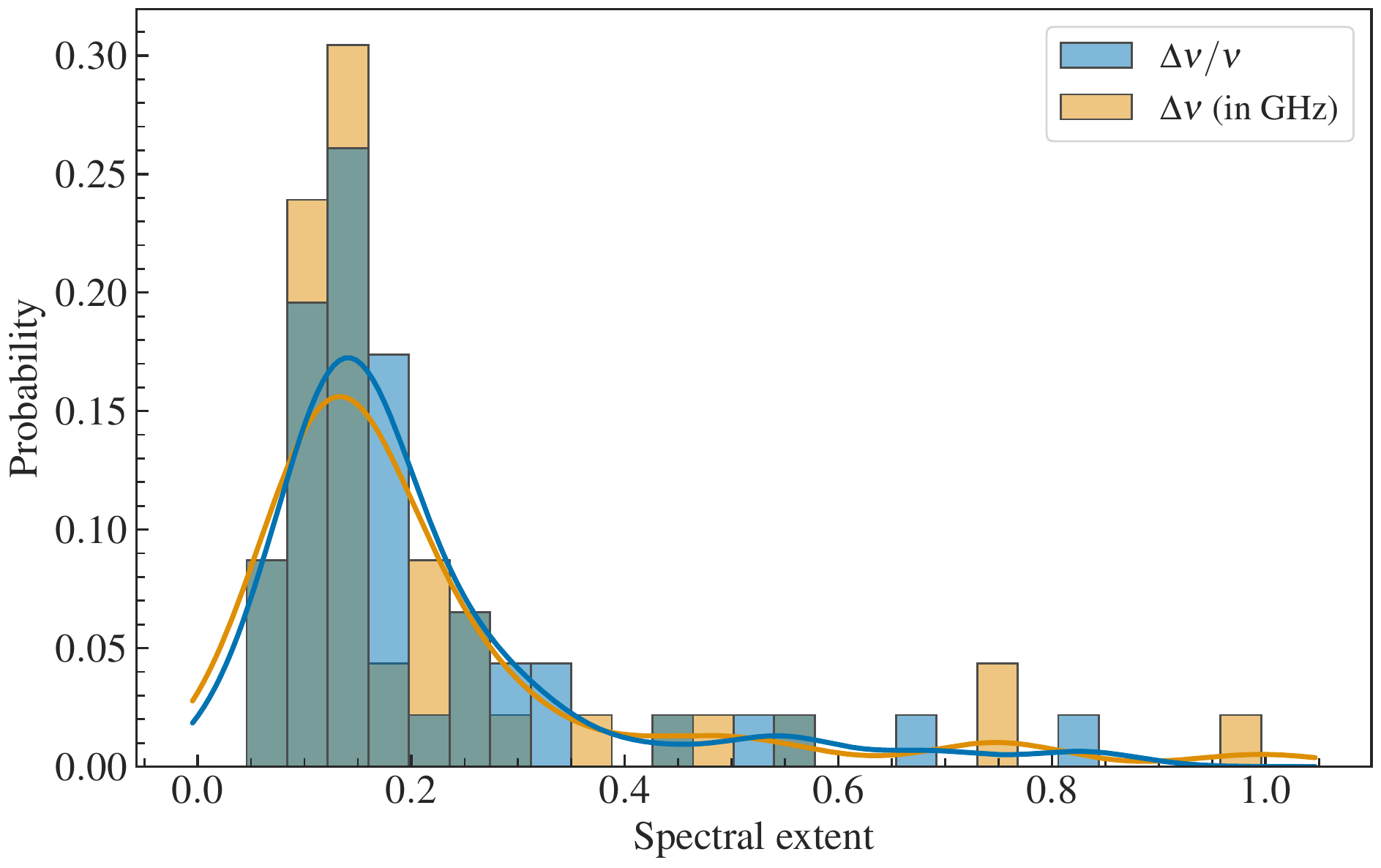} 
\caption{Distribution of the spectral extents of the Parkes-detected repeat bursts from the FRB\,20180301A source. The spectral extents are represented as the fractional bandwidth ($\Delta \nu / \nu$) and absolute bandwidth ($\Delta \nu$). The respective distributions are depicted using Gaussian kernel density estimates.}
\label{fig:spectral_extent}
\end{figure}

We use the respective group average DMs to measure the burst properties reported in Table \ref{tab:burst_properties} and elsewhere in the paper. We find that the obtained group DMs adequately explain the frequency-time structure for most bursts of their respective groups. However, a burst-to-burst DM variation cannot be ignored. Notably, in the bursts with the narrowest temporal width in our sample $\sim$100\,$\upmu$s (B10 and B16), the extra DM $\sim$0.5\DMunits is clearly apparent from the burst structure (see Fig.~\ref{fig:repeaterplots}). They might also contain sub-$\upmu$s level burst components, unresolved given our limited time resolution of 64\,$\upmu$s. Due to their extremely narrow pulse width, this discrepancy becomes quite significant when measuring burst properties, as any slight deviation distorts the burst structure (dispersion delay across spectral envelope $>>$ pulse width). So, for these two bursts, we use their individual best DM$_\mathrm{struct}$ to measure burst properties.

Measuring the DM of bursts using a structure-maximizing algorithm can be complicated when the burst signal contain irregular or complex morphologies. For instance, burst B37 exhibits an unusual and irregular signal structure that is qualitatively similar to the burst morphologies observed in some bursts from FRB\,20201124A \citep{Zhou:2022}. The signal consists of two sub-components with a spectral extent of 704--900 and 900--1400~MHz, respectively. The high-frequency component drifts downwards, whereas the low-frequency one shows no such feature. Undoubtedly, both components are morphologically distinct. We must ascertain a proper DM to interpret the burst signal structure reasonably. The DM$_\mathrm{struct}$ is $518.9\pm0.3$\DMunits for the entire burst and $519.0\pm0.3$\DMunits for the high-frequency component. The low-frequency component has a straightforward signal structure, and the S/N maximisation provides DM$= 514.7\pm0.8$\DMunits. Thus, we are presented with two DM scenarios, each trying to align the individual sub-components in time. Thus, a single DM can not be determined using existing methods to explain the entirety of the burst structure. For these reasons, we exclude B37 in determining the group average DM. The de-dispersed burst spectrum and the pulse profiles for both extreme DM scenarios are shown in Fig.~\ref{fig:dm_comp}. We confirm that the group average DM better represents the signal structure in this repeat burst.
 
\subsection{Burst properties}
We do not find evidence of radio signal emission above 1.8\,GHz in any of the Parkes-detected repeat bursts from FRB\,20180301A source. The bursts are predominantly centred in the lower part, but not at the bottom,  of the UWL band. We extract the portion of the dynamic spectrum where the burst signal is present to characterise its properties. We estimate the spectral envelope by visual inspection, ensuring that it enclosed all of the burst signals. The shaded region in Fig.~\ref{fig:repeaterplots} shows the spectral envelope ($\nu_\mathrm{low}$ and $\nu_\mathrm{high}$) that was used to measure the burst properties, and the values are reported in Table~\ref{tab:burst_properties}. Because the majority of UWL bursts occur in the lower band, which is heavily affected by RFI, a significant portion of the frequency channels (up to $\sim50$ per cent) are excised to mitigate their effects. As a result, determining a clear edge of the spectral envelope in many UWL repeat bursts is challenging. We use the centre of the adjacent excised band as the spectral edge for those bursts, and the zapped channel bandwidth as its associated uncertainty. Furthermore, the signal spectral envelope extends to the lower end of the UWL bandwidth in $\approx 30$ per cent of the repeat bursts (see Table~\ref{tab:burst_properties}). Because RFI corrupts the bottom-most frequency channels (< 750\,MHz), it is difficult to ascertain the lower extent of the bursts. So, for these bursts, we conservatively assume the lowest UWL frequency channel $= 704$\,MHz as the edge of the burst spectral envelope. In addition, the sensitivity over the UWL band may vary significantly, particularly at lower frequencies (< 1\,GHz), which could have affected the burst detection rate at these frequencies.

In comparison to the total extent of the UWL band, all Parkes-detected repeat bursts are band-limited. The average central emission frequency of bursts is estimated to be 1.1\,GHz. The average burst spectral occupancy is $\sim 220$~MHz, which accounts for only $\approx 7$ per cent of the total observing UWL band. This estimate is also a lower limit, given the uncertainty in the lower frequency end of the spectral envelope for several bursts. We quantify the spectral extent of the bursts using the absolute spectral extent $\Delta \nu$ as well as the fractional extent at the burst central emission frequency $\Delta \nu / \nu$. The spectral extent distribution of the Parkes/UWL sample is shown in Fig.~\ref{fig:spectral_extent}. The repeat bursts exhibit a wide range of fractional bandwidth with $\Delta \nu / \nu$ ranging from 0.05 to 0.83. A few of these repeat bursts, which are coincidentally also relatively bright in the UWL sample (B37--B39) have broad spectral extent in the range 0.7--1.0\,GHz, corresponding to a fractional bandwidth of 0.54--0.83. 

We use least-square minimization to measure the temporal width of the repeat bursts, fitting the frequency-averaged pulse profile with a single Gaussian function. We conduct the fitting at a time resolution of $0.5$--$1$\,ms, except for the narrow bursts B10 and B16 for which we use the full temporal resolution data. To determine the integrated pulse S/N, we convolve the pulse profile with a set of Square-normalised Gaussian templates, of varying widths, using the package \software{spyden\footnote{\url{https://bitbucket.org/vmorello/spyden}}}. The S/N value of the best-fitting template is reported in Table~\ref{tab:burst_properties}. For bursts with multiple sub-components, we, however, make use of a boxcar template. To determine burst fluence, we integrate over twice the measured Gaussian full width at half-maximum (FWHM) duration of the burst as the on-pulse region to include all of the burst power. Except for the two unusually narrow bursts $\sim100\upmu$s (B10 and B16) and some other bursts with multiple components, most bursts in our sample have temporal width ranging from  2 to 6\,ms. For broader bursts, the data were decimated in time before measuring their properties in order to be consistent with single peak bursts. 

We do not find strong morphological evidence for temporal scattering in single-peaked repeat bursts. The NE2001 and YMW16 galactic models predict a scattering time-scale of 10\,$\upmu$s and 1\,ms respectively at 1\,GHz along the line of sight of this FRB source \citep{Cordes:2002, Yao:2017, Price:2021}. B10 and B16 are the narrowest bursts in the UWL sample with high S/N. These two bursts also show single-peaked pulse profiles, which provides an opportunity to constrain scattering timescales. We adopt a Bayesian approach to model the pulse profiles to investigate the possibility of pulse broadening caused by multi-path scattering \citep{Ravi:2019_properties}. We convolve an exponential decay function with a Gaussian base model to create a broadened burst profile. We construct the model for a sampling interval ten times higher than the actual burst profile and then average it to the final temporal resolution to account for any unresolved components \citep{Qiu:2020}. We then compute the ratio of Bayesian evidence ($\mathcal{B}$) between models with and without scattering to determine evidence for scattering. For burst B10, we find that a scattered profile is strongly favoured with $\log_{10}\mathcal{B} > 10$ \citep{Trotta:2008}. Modelling with a scattered profile results in a significantly narrower pulse width (FWHM) of $55_{-26}^{+18}\,\upmu$s with a scattering time-scale $\tau_{\rm s}$ of $156\pm13\,\upmu$s at 1.12\,GHz. For burst B16, we do not find evidence for scattering and limit $\tau_{\rm s}$ to be $<26\,\upmu$s at 1.7\,GHz. We note that this upper limit is consistent with a frequency evolution of $\tau_{\rm s} \propto \nu^{-4}$ expected from a thin scattering screen \citep{Scheuer:1968}. Establishing scattering in bursts with multiple peaks or frequency-drifting is difficult, so we do not attempt to model scattering in multi component bursts.

\begin{figure}
\centering
\includegraphics[width=\columnwidth]{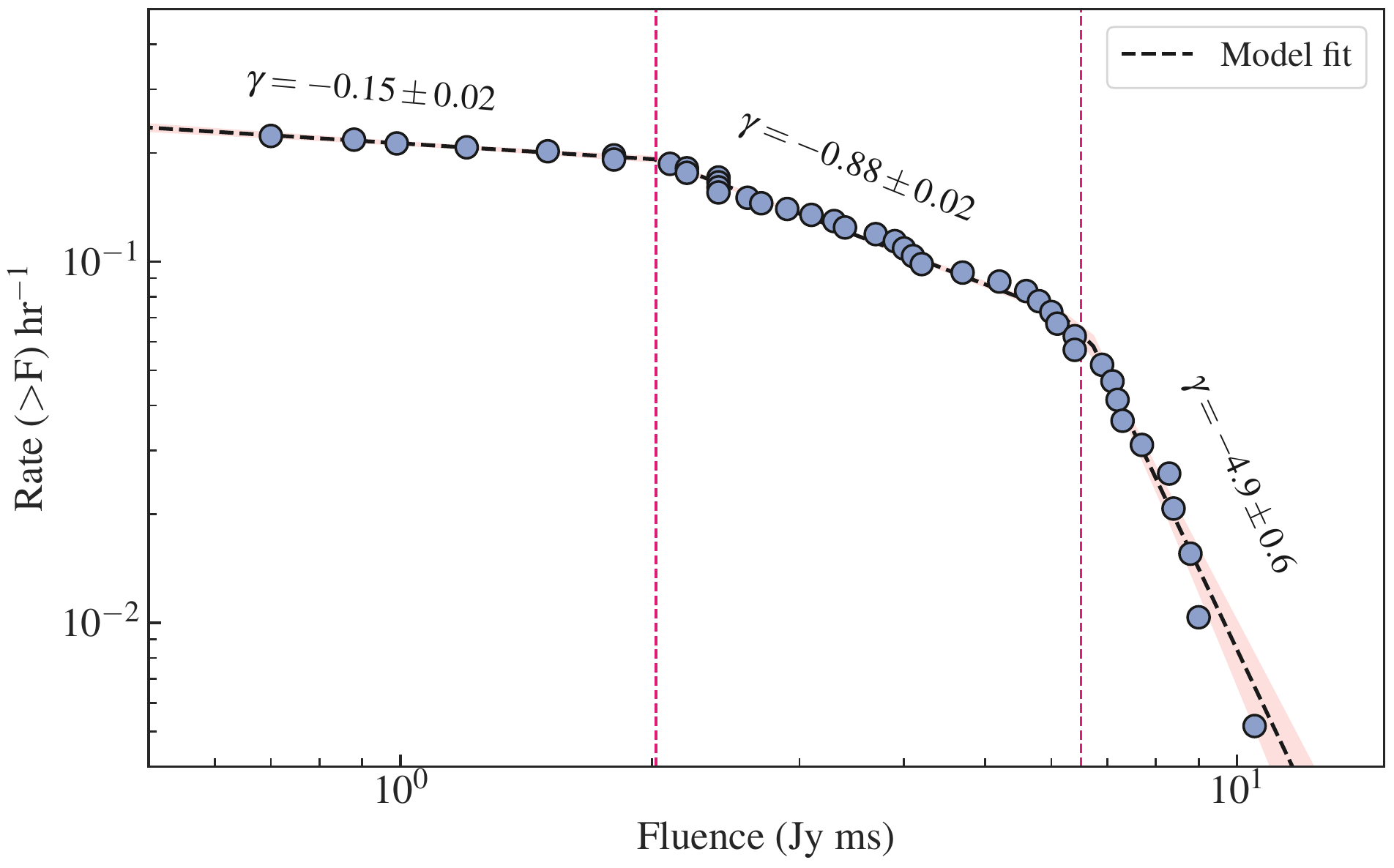}
\caption{Cumulative burst rate function of the Parkes-detected repeat bursts. The red dashed lines indicate the break fluences obtained from a fit with a broken power-law model. The black dashed line shows the best-fitting model with the respective power-law indices $\gamma$ labelled on the top.}
\label{fig:fluence_cdf}
\end{figure}

\begin{figure}
\centering
\includegraphics[width=\columnwidth]{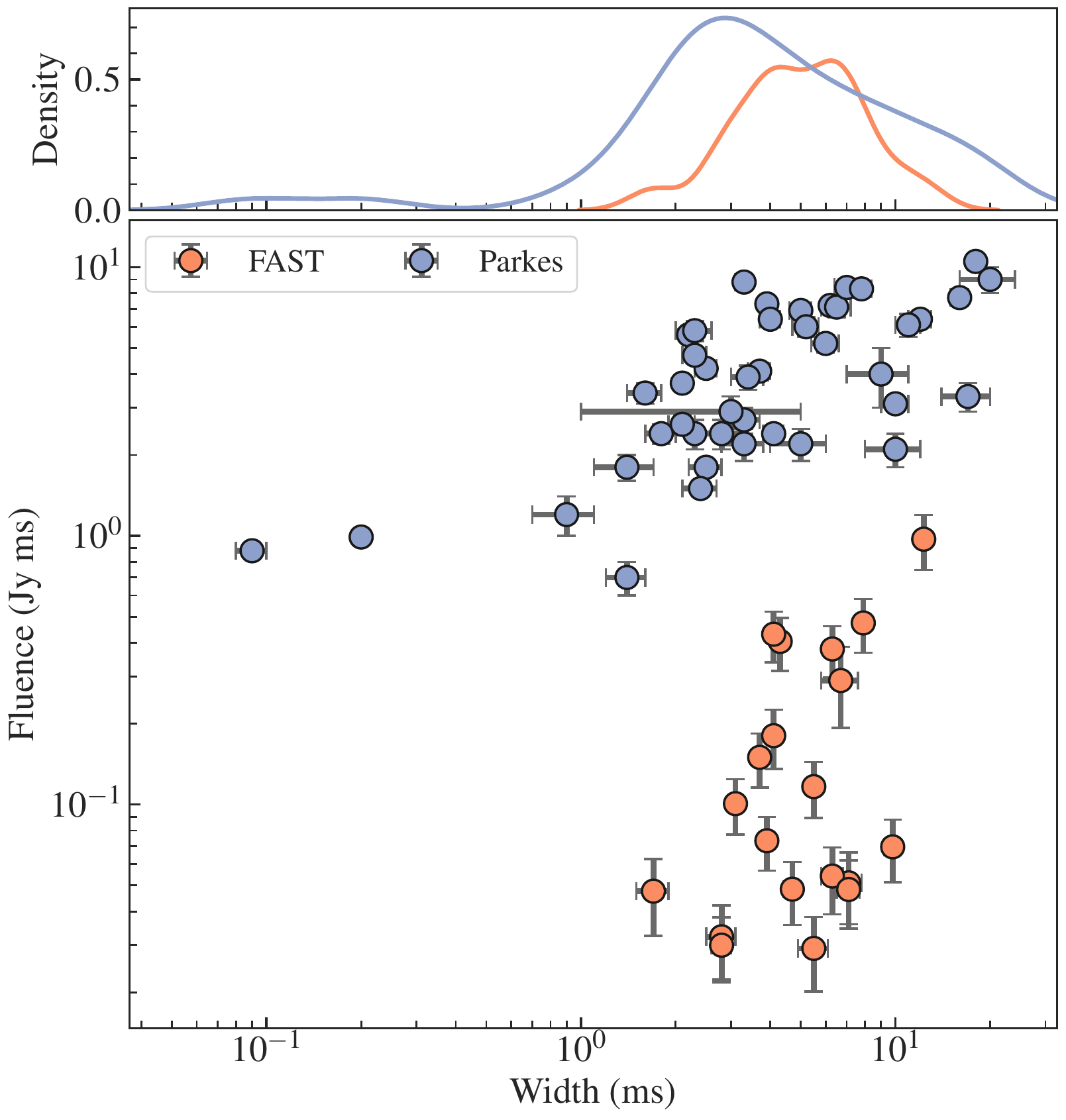} 
\caption{Fluence-width distribution of Parkes-detected repeat bursts from FRB\,20180301A source. The FAST-detected bursts are also shown for comparison \citep{Luo:2020, Laha:2022}. The top panel displays the Gaussian kernel density estimate of the width distribution for both burst samples.}
\label{fig:fluence_width}
\end{figure}

We calculate the cumulative distribution function (CDF) of the Parkes burst rate above their detected fluences. The CDF has multiple power law features with breaks at low and high fluences. We model the CDF using a broken power law function of the form $R(>F) \propto F{\nu}^{\gamma}$, in which $R$ is the rate (in h$^{-1}$), $F$ the fluence (in Jy ms) and $\gamma$ the power-law index. We utilize a least-square minimization technique and a broken power law with two turn-overs to find that the break fluences are located at $6.5\pm0.1$ and $2.0\pm0.1$\,Jy\,ms. The best-fitting power law has index $\gamma = -0.15\pm0.02, -0.88\pm0.02, -4.9\pm0.6$ in the regions respectively. Figure~\ref{fig:fluence_cdf} shows the fluence CDF and the best-fitting model for Parkes-detected repeat bursts. The lower turn-over can likely be attributed to the incompleteness of the sub-band searches for bursts close to the sensitivity limits presented in Table~\ref{tab:subband_params}. We attribute the break fluence $\sim$2\,Jy\,ms to the completeness limit of Parkes observations. We only attempt to obtain an empirical fit to the CDF to estimate the completeness limits of the search techniques and UWL receiver. A more rigorous analysis of fluence completeness is beyond the scope of this paper. We present a comparison of the measured fluence-width distribution for the Parkes-detected repeat bursts with the FAST sample in Fig.~\ref{fig:fluence_width}. 

B38 is the widest burst in our sample, both spatially and temporally. It exhibits a wedge-like structure extending from $\sim$1700\,MHz to the bottom of the UWL band. The majority of the burst signal is concentrated in the central sub-structure, with the top and bottom contributing significantly less to the S/N. As a result, determining the actual temporal extent of the burst, which appears to be $\sim$30\,ms upon visual inspection, is challenging. To better analyse the burst, we divide the structure into three constituents with spectral extents of 704--1004, 1004--1404, and 1404--1700\,MHz, respectively. In addition to analyzing the entire spectral extent, we also provide burst properties for B38 using only the central component in Table~\ref{tab:burst_properties}. 

\begin{figure*}
\centering
\includegraphics[width=\textwidth]{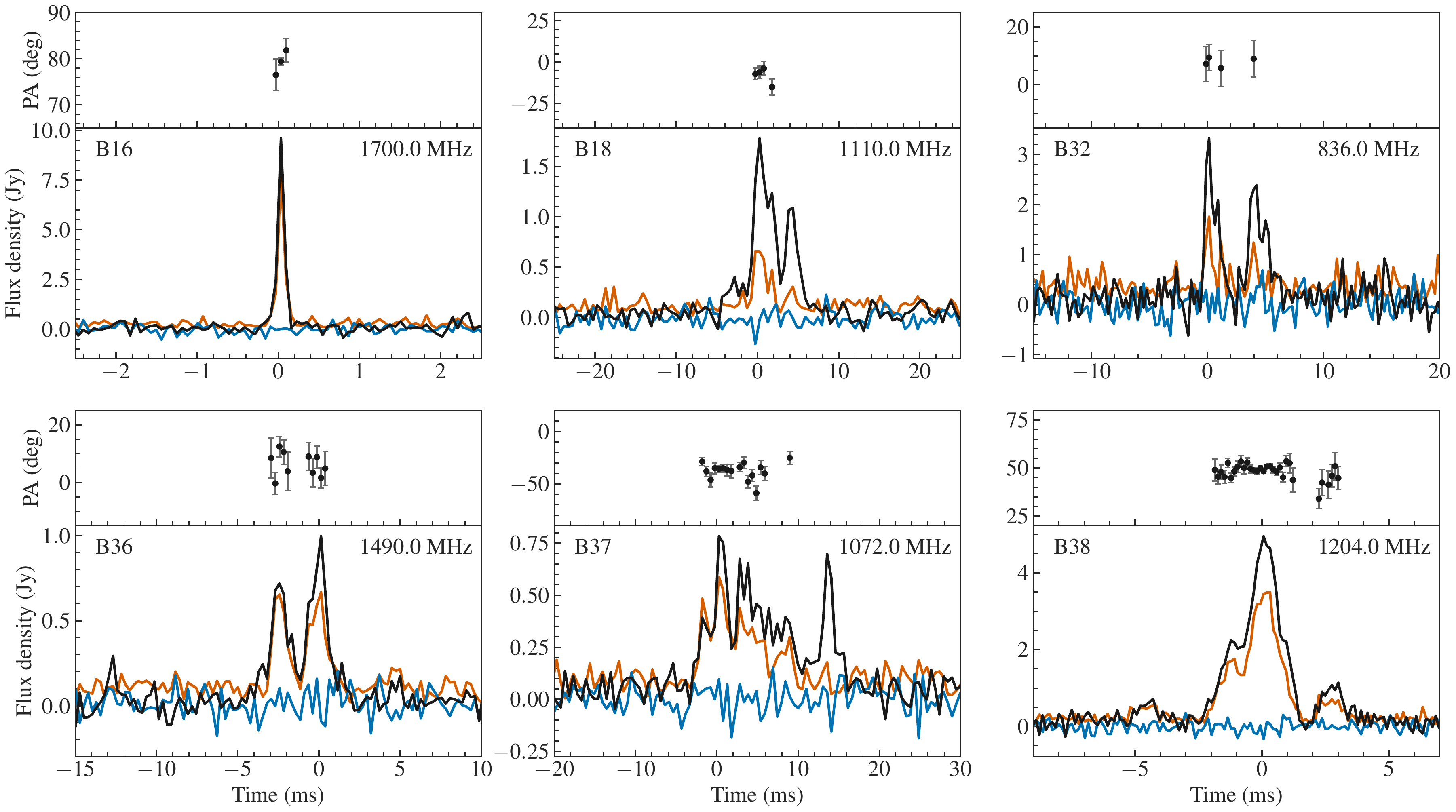} 
\caption{Polarization profile of a selection of Parkes-detected repeat bursts from the FRB\,20180301A source. In each panel, the top sub-plot displays the de-biased linear polarization PA. The bottom sub-plot shows the frequency-averaged time series for Stokes I (black), V (blue) and total linear polarization, L (orange). The burst data are corrected for the individual best-fitting RM using the centre of the burst sub-band at the reference frequency (labelled in top right).}
\label{fig:polprofile}
\end{figure*}

\subsection{Polarimetry}
More than two-thirds of the Parkes-detected repeat bursts show no evidence for a detectable polarized flux in their Stokes spectra. Since many of these bursts have low S/N ($<15$), only weak constraints can be placed on the level of polarized flux. Nevertheless, this does not account for the lack of polarized flux in bursts with moderate to high S/N, such as B14--B15, B19--B21 and B23. A large Faraday RM could have caused the depolarization of the Stokes $Q$ and $U$ signal across the coarse frequency channels \citep{Michilli:2018}. In terms of the PA, the Faraday rotation of a radio signal within a single frequency channel is given as \citep{Burke:2019}
\begin{equation}
\Delta \theta = \frac{(1.8 \times 10^5)\,\delta\nu\,\text{RM}}{\nu_{c}^3} \text{rad},
\end{equation}
where $\delta\nu$ is the channel resolution in MHz at a central channel frequency $\nu_{c}$. The variations in PA within a coarse frequency channel cause LP to leak from $Q$ into $U$ and change sign, which will result in depolarization and an underestimation of LP fractional strength. For this intra-channel depolarization, the fractional reduction in the linear polarization (LP) is commonly expressed in terms of the ratio of the respective PA magnitudes as \citep{Michilli:2018}
\begin{equation}
f_{\text{depol}} = 1 - \left| \frac{\text{sin}(\Delta \theta)}{\Delta \theta} \right|,
\end{equation}
This effect is more pronounced at lower frequencies and can completely depolarize a burst signal. Assuming we can detect $\gtrsim$20 per cent linearly polarized signal (3$\sigma$ limit for S/N = 15), it would require the intra-channel rotation $\Delta \theta \gtrsim$ 1 rad. Given the UWL channel bandwidth of $\delta\nu = 0.5$\,MHz and the average central emission frequency of $\nu = 1.1$\,GHz, we obtain a lower limit on the Faraday RM that can depolarize the signal to be $|\mathrm{RM_{depol}}| = 1.5 \times 10^4$\RMunits. Taking this as the maximum detectable RM, we use the \software{rmfit} tool from \software{psrchive} to perform a brute-force RM search on the UWL repeat bursts with no visible polarized signal. We search for a range of trial RMs between $\pm \mathrm{RM_{depol}}$ with a step size of 1 \RMunits. We did not identify any peak in the polarized intensity for the RM trials used in the search for all those bursts. We note that the brute-force RM search does not correct for any bandwidth depolarization which might have resulted in non-detections. These bursts and their spectral coverage are indicated in the timeline of repeat burst observations in Fig.~\ref{fig:timeline}.

Of the 46 UWL-detected repeat bursts from FRB\,20180301A, 13 had detectable polarized emission. These bursts are in two sets and separated by a year. They are B10 and B16--B18 in group G1 and B32--B39 in G2. We use \software{RMNest} \citep{Lower:2022}, which utilizes Bayesian parameter estimation to infer the Faraday RM$_\mathrm{nest}$ for these bursts by directly fitting the Stokes Q and U spectra. We use a uniform prior on RM in the range $\pm \mathrm{RM_{depol}}$. We also use RM synthesis \citep{Brentjens:2005} to form the Faraday dispersion function (FDF) through a Fourier transform of the Stokes Q and U spectra. We use the \software{rmsynth1d} from the \software{RM-TOOLS} package \citep{Purcell:2020} and search for 20000 trial RMs in the range $\pm \mathrm{RM_{depol}}$. The resulting RM measurements with 1$\sigma$ uncertainties using both methods are presented in Table \ref{tab:rm_table}. We find an extensive variation in the measured RM values of these bursts over a period of two years. The Faraday RM$_\mathrm{nest}$ of bursts varies around their weighted mean value of $-82$\RMunits with a standard deviation of $59$\RMunits. Not only these are distinct from the RM measurements $\sim542$\RMunits reported earlier with the FAST-detected bursts \citep{Luo:2020}, but the sign of Faraday RM has also reversed. The significant RM variation also adds credibility to the RM$=-3163$\RMunits measured for the original FRB\,20180301A detection \citep{Price:2019}. It also suggests that the FRB progenitor has gone through multiple RM sign-reversals in recent years.

\begin{table}
\caption{Faraday rotation measure of UWL repeat bursts obtained with \software{rmsynth1d} and direct Stokes Q--U fitting with \software{RMNest}. Polarization fractions are obtained after correcting for individual RM$_\mathrm{nest}$. Uncertainties are given as 1$\sigma$ confidence interval.} 
\label{tab:rm_table}
\normalsize
\centering
\begin{tabulary}{\columnwidth}{LRRRRR}
\hline \hline
Burst & RM$_\mathrm{synth}$ & RM$_\mathrm{nest}$ & P/I & L/I & V/I \\
      &(\RMunits) & (\RMunits) &  & & \\
\hline
B10 & $-$124(8) & $-$122(10) & 0.50(5) & 0.50(5) & 0.01(4)\\
B11 & $-$210(65)& $-$212(40) & 0.8(2)  & 0.8(2)  & 0.1(1)\\
B16 & $-$154(5) & $-$157(17) & 0.80(3) & 0.80(3) & 0.01(3)\\
B17 & $-$237(31)& $-$184(45) & 0.63(9) & 0.59(9) & $-$0.04(9)\\
B18 & $-$237(5) & $-$232(7)  & 0.33(3) & 0.33(3) & $-$0.03(3)\\
B32 & $-$168(10)& $-$172(14) & 0.42(5) & 0.38(3) & 0.02(5)\\
B33 & $-$149(36)& $-$135(28) & 0.9(1)  & 0.8(1)  & $-$0.1(1)\\
B34 & $-$3(3)   & $-$5(3)    & 0.67(5) & 0.67(5) & $-$0.03(4)\\
B35 & $-$123(30)& $-$86(25)  & 0.6(1)  & 0.6(1)  & $-$0.1(1)\\
B36 & $-$104(5) & $-$103(4)  & 0.75(4) & 0.74(4) & 0.03(3)\\
B37 & $-$94(1)  & $-$92(2)   & 0.74(4) & 0.73(4) & 0.01(3)\\
B38 & $-$87(2)  & $-$82(2)   & 0.72(1) & 0.71(1) & $-$0.01(3)\\
B39 & $-$98(7)  & $-$107(9)  & 0.65(3) & 0.61(3) & 0.15(4)\\
\hline
\end{tabulary}
\end{table}

The Faraday RM of the Parkes-detected repeat bursts shows significant temporal variations. We use a linear model to fit for variations in RM$_\mathrm{nest}$ and obtain a gradient of
\begin{equation}
    \frac{d\rm RM}{dt} = +98\pm55\,{\rm \,rad\,m^{-2}\,yr^{-1}}.
\end{equation}
The fit to the RM variation is inadequate given the large uncertainties associated with the measured RMs, but it undoubtedly captures the trend. Similar to the classification scheme we used earlier for burst DM, we define a global RM for the FRB source as the weighted mean of RM$_\mathrm{nest}$ measured for bursts in the respective groups. We obtain the group RM to be $-192\pm 25$ \RMunits for bursts in G1 and $-68 \pm 16$ \RMunits for G2. The obtained RM slope indicates that the average FRB source RM for G2 bursts has increased by 64 per cent with more than $4\sigma$ significance compared to G1 bursts. Considering the significant variation in burst-to-burst RM, we use the individual burst RMs throughout this paper to correct their Stokes spectra and measure the polarization strength.

We obtain the polarimetric pulse profile for the polarized UWL repeat bursts by averaging the Faraday-corrected Stokes spectra over frequency. We use the RM$_\mathrm{nest}$ values to correct for Faraday rotation in all 13 repeat bursts. For each burst, we calculate the absolute linear polarization PA using the frequency-averaged Stokes $Q$ and $U$. We remove the bias in the total linear polarization $L$ for each time sample using \citep{Everett:2001, Day:2020}
\begin{equation}
L_\mathrm{de-bias} = \begin{cases}
\sigma_I \sqrt{\left( \frac{L_\mathrm{meas}}{\sigma_I} \right)^2 - 1} & \mathrm{if~} \frac{L_\mathrm{meas}}{\sigma_I} \geq 1.57 \\
0 & \mathrm{otherwise,}
\end{cases}
\end{equation}
where $\sigma_I$ is the Stokes $I$ off-pulse standard deviation and $L_\mathrm{meas} = \sqrt{Q^2 + U^2}$. We use a 3.5$\sigma$ threshold on the de-biased $L$ to establish the significant measurements of PA. The frequency-averaged Stokes $I$, $L$, and $V$ profile of a sample of the polarized repeat bursts, along with the significant measurements of PA, are plotted in Fig.~\ref{fig:polprofile}. No clear evidence of a significant variation in the PA profile is observed in the UWL bursts. This is in contrast to the PA variations observed in bursts detected from this source with FAST \citep{Luo:2020}. We also measure the total $P/I$, the linear $L/I$, and the circular polarization fractions $V/I$ for each burst, where $P = \sqrt{L_\mathrm{de-bias}^2 + V^2}$ is the total polarization. Uncertainties are calculated based on the principles of error propagation from the off-pulse standard deviation in the Stokes spectra. The polarization fractions with their uncertainties are presented in Table~\ref{tab:rm_table}. We do not find evidence of circular polarization in any of the 13 polarized UWL bursts. The amount of fractional LP is not constant, unlike other repeating FRBs, but has a significant variation in the range of 30--80 per cent. 

In the limited sample of polarized bursts, a trend of decreasing degrees of LP with decreasing frequencies is apparent. The diminished fraction of LP at lower frequencies cannot be explained solely by intra-channel depolarization. A possible explanation for the the depolarization is RM scattering \citep{Sullivan:2012}. Multipath radio signal transmission in an inhomogeneous magneto-ionic environment could cause such scattering. The fractional reduction in the LP is parameterized as \citep{Feng:2022}
\begin{equation}
f_{\text{RM scattering}} = 1 - \exp(-2\,\lambda^4\sigma_{\rm RM}^2),
\label{eq:rm_scatter}
\end{equation}
where $\sigma_{\rm RM}$ is the standard deviation in the RM and $\lambda$ is the central emission wavelength of the burst.

\begin{figure}
\centering
\includegraphics[width=\columnwidth]{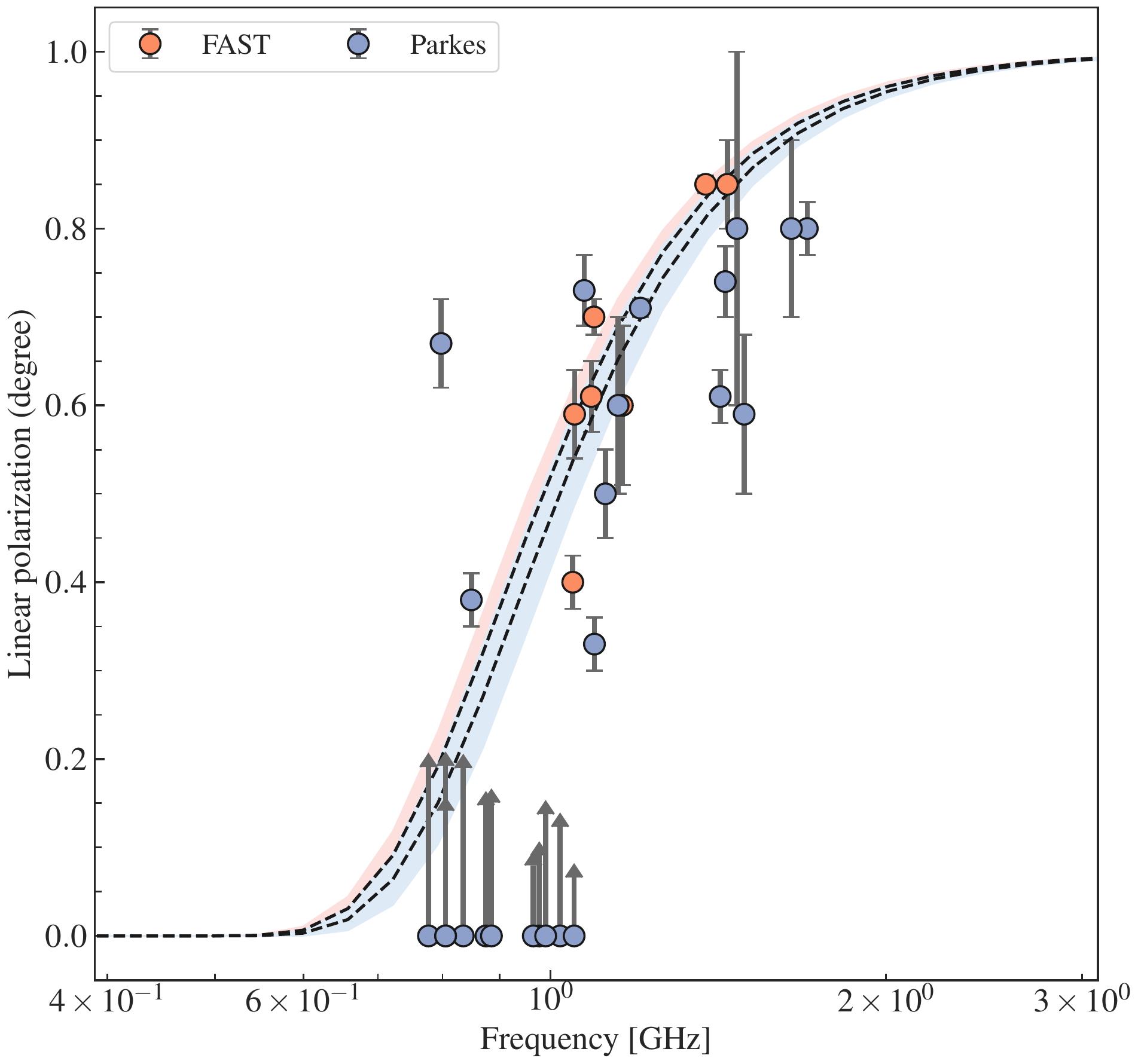} 
\caption{Depolarization of Parkes-detected bursts as a function of observing frequency. The FAST-detected bursts are also shown for comparison \citep{Luo:2020, Feng:2022}. The best-fit model using Equation~\ref{eq:rm_scatter}, represented by black dashed lines, along with the 1$\sigma$ uncertainty shaded region, is displayed. Additionally, 3$\sigma$ upper limits for high S/N (>15) unpolarized Parkes-detected bursts are also plotted.}
\label{fig:burst_depol}
\end{figure}

We determine $\sigma_{\rm RM}$ for the Parkes-detected polarized bursts by fitting the measured degree of LP with the above model using a least-squares minimization fit. We assume that each repeat burst is intrinsically 100 per cent linearly polarized. The central emission frequency of the bursts is determined by taking the mean of the burst spectral envelope (reported in Table~\ref{tab:burst_properties}). There are few high frequency bursts (B08 and B27) in our sample without detectable polarization and several low frequency bursts with polarization, implying a scatter in $\sigma_{\rm RM}$. We obtain $\sigma_{\rm RM} = 6.8\pm0.5$\RMunits for UWL-detected repeat bursts. For the FAST-detected repeat bursts, \citet{Feng:2022} reported  $\sigma_{\rm RM} = 6.3\pm0.4$\RMunits. We also include the FAST measurements to do a combined fit for all the repeat bursts and obtain $\sigma_{\rm RM} = 6.6\pm0.4$\RMunits. All the above inferred $\sigma_{\rm RM}$ values are reported in the observer's frame. In Fig.~\ref{fig:burst_depol}, we present the degree of LP versus frequency for both FAST and UWL repeat bursts, including their respective model fits. We also include high S/N (>15) unpolarized Parkes-detected bursts with their 3$\sigma$ upper limit in the plot.

\begin{table}
\caption{Sub-pulse drift rates for the Parkes-detected repeat bursts obtained using a 2D Gaussian fit to the ACF of the intensity dynamic spectrum. The temporal width and spectral band extents are defined as $1/\sqrt{2}$ multiplied by the FWHM of the Gaussian fit. The uncertainties are not provided, as they cannot be well quantified from the Gaussian fit.} 
\label{tab:drift_table}
\normalsize
\centering
\begin{tabulary}{\columnwidth}{LRRRR}
\hline \hline
Burst ID & $\nu_\mathrm{centre}$ & Width &  Band extent & Drift rate \\
         & (MHz)  & (ms)  & (MHz)   & (\driftunits)\\
\hline
B14 &    1020.0 & 6.8  &  245.8 &$-$34.2\\
B15 &     977.0 & 10.5 &  317.8 &$-$19.3\\
B18 &    1095.0 & 5.6  &  214.3 &$-$48.7\\
B34 &     797.5 & 7.8  &  114.1 &$-$20.6\\
B36 &    1435.0 & 3.3  &  310.2 &$-$100.7\\
B37A&    1140.0 & 9.5  &  348.1 &$-$46.5\\
B38 &    1202.0 & 5.6  &  248.9 &$-$61.7\\
B39 &    1420.0 & 6.3  &  429.0 &$-$42.4\\
\hline
\end{tabulary}
\end{table}

\subsection{Sub-pulse drifts}
The Parkes-detected sample of repeat bursts from the FRB 20180301A source contains a wide range of burst morphology. A complex multi-component signal structure can be observed in several of these bursts. Eight of them (B14, B15, B18, B34, and B36--B39) have high S/N and show evidence of sub-pulses drifting with frequency. This characteristic drifting property has previously been observed in bursts from other repeating FRB sources \citep{CHIME:2019_8repeaters, Hessels:2019, Zhou:2022}. We also find that, like them, the sub-bursts drift towards lower frequencies at later times in the burst envelope (the ``sad trombone'' effect). We find no evidence of upward drifting in any of the repeat bursts in our sample. We estimate the linear drift rates for these repeat bursts using the standard approach of computing the 2D autocorrelation function (ACF) for the on-pulse intensity dynamic spectrum. Before computing the ACF, we dedisperse the signal structure using each burst individual DM$_\mathrm{struct}$. The zero-lag noise spike in the computed ACF is then removed in both time and frequency axis. By fitting a generalized 2D Gaussian to the ACF, the linear drift rate, $\dot{\nu} = d\nu /dt$ defined as the tilt in the ACF ellipse in units of \driftunits is measured \citep{Hessels:2019}. Table \ref{tab:drift_table} displays the obtained sub-pulse drift rates for the repeat bursts. The uncertainties associated with the drift rate are not well quantifiable for the direct Gaussian fitting approach, so we do not include them here.

\begin{figure}
\centering
\includegraphics[width=\columnwidth]{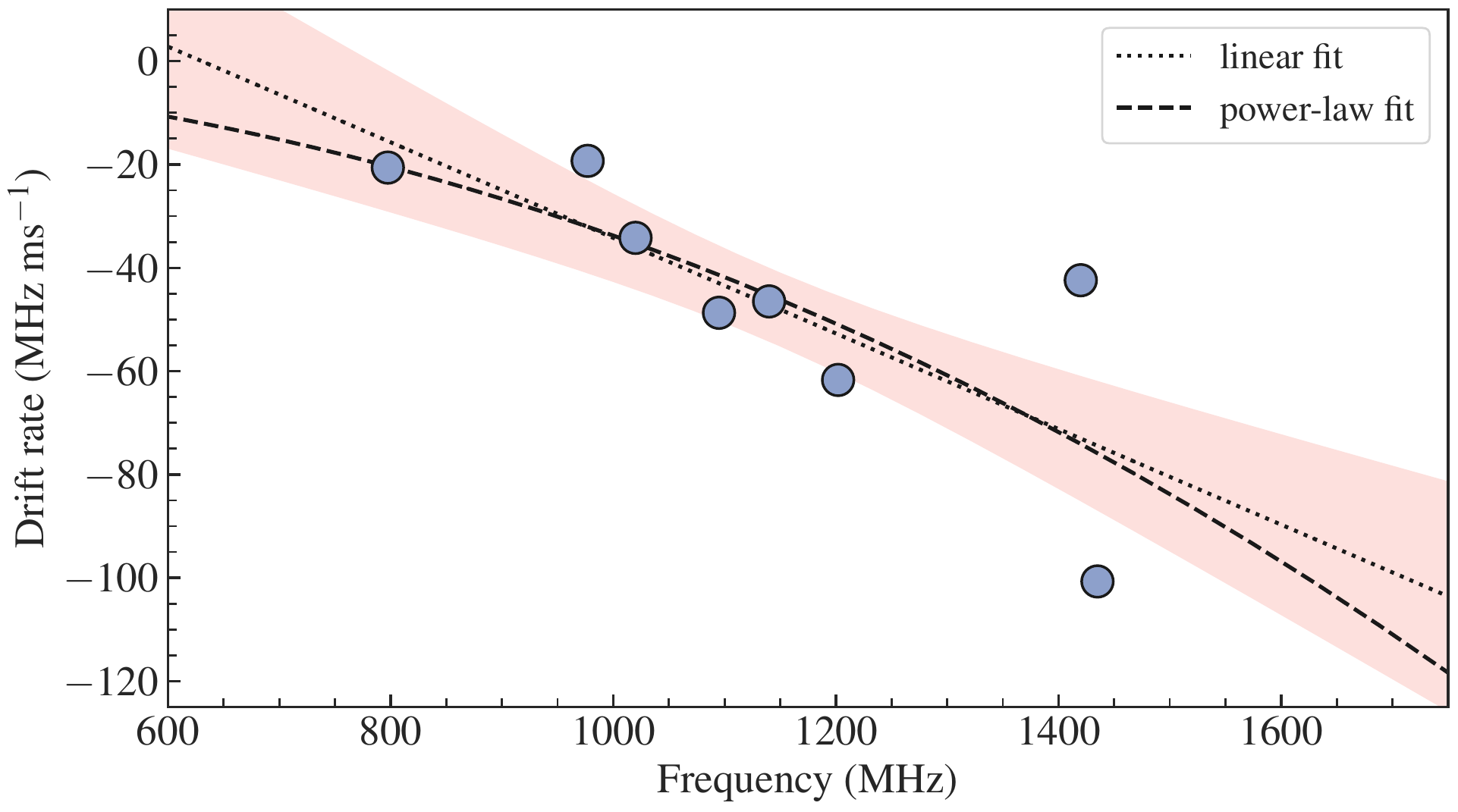} 
\caption{Spectro-temporal linear drift rates as a function of central observing frequencies for Parkes-detected repeat bursts from FRB\,20180301A. A least-squares fit for a linear regression model provides $\dot{\nu} \propto -0.09\nu$, indicating a clear increase in the rate magnitude with observing frequency. The dotted black line represents the best-fitting linear model, with the shaded background indicating the 1$\sigma$ uncertainty. The power-law fit is shown as a dashed line.}
\label{fig:drift_rates}
\end{figure}

We obtain an average sub-pulse drift rate of $-48\pm11$\driftunits at an average observing frequency of 1136\,MHz. We find that the magnitude of the drift rate increases with increasing radio frequency, which is consistent with observations of other repeating FRB sources \citep{Marazuela:2021, Platts:2021}. To quantify the drift rate evolution with frequency, we fitted a linear regression model and obtained the best-fitting slope $=-0.09\pm0.03$\,ms$^{-1}$. We also attempt to model the rate evolution using a power-law function and obtain the power-law index $=2.2\pm0.8$. The measured drift rates for the Parkes-detected repeat bursts and their best-fitting models are shown in Fig.~\ref{fig:drift_rates}. A 1D Gaussian function is insufficient to describe FRB temporal profiles with complex sub-structures. We additionally utilise the computed 2D ACF to calculate the burst extent in time and frequency. We do not use the same approach for all repeat bursts as the low S/N and other morphological effects (such as scintillation) could result in an underestimation of the burst extents. We define the FWHM of the temporal envelope and the spectral band extents as $1/\sqrt{2}$ multiplied by the FWHM obtained from the 2D Gaussian fit ACF to the dynamic spectrum \citep{Nimmo:2022}. For these bursts, we obtain an average temporal extent $\sim$7\,ms and a spectral band extent $\sim$279\,MHz. The burst widths and the drift rates determined by the 2D ACF are also presented in Table \ref{tab:drift_table}.

Not all burst morphology in our sample demonstrates clear evidence of sub-pulse drifting. The pulse envelope in repeat bursts B19, B23, and B32 contains multiple components, though their individual spectral extent appears to be comparable. It is difficult to tell whether these sub-components are drifting in frequency or are entirely separate bursts separated in time. We do not use the above analysis for these bursts because it would have resulted in large negative drift rates given the small tilt angle values of the ACF ellipse \citep{Pleunis:2020PhDT}. A straightforward definition for burst signal boundaries is an ongoing area of investigation, and whether to call them multiple sub-bursts inside one burst envelope or multiple separate burst envelopes without a link in emission is unclear at this stage \citep{Pleunis:2021_LOFAR, Jahns:2022}.

\subsection{Spectral properties}
Radio spectral analysis is a critical tool for understanding the emission mechanisms of astrophysical sources, including FRBs. An accurate representation of repeating FRBs spectra has yet to be realized owing to the unavailability of wideband (> GHz) spectral data. Additionally, the narrow spectral extent of these bursts presents further challenges to spectral modelling and gaining an intuition for the source spectra \citep{Agarwal:2020}. It remains unclear whether the narrow-band spectra are intrinsically produced in the FRB progenitor or are hindered/amplified by propagation effects along the line of sight \citep{Cordes:2019}. Unlike FRBs, pulsar spectra typically span a wider frequency range and can be easily modelled using a single power-law function or multiple power-law functions (with breaks) considering spectral turnover \citep{Aggarwal:2022}. However, a power-law model has not been successful in modelling individual FRB spectra, and several alternative models have been proposed \citep{Pleunis:2021}.

\begin{figure}
\centering
\includegraphics[width=\columnwidth]{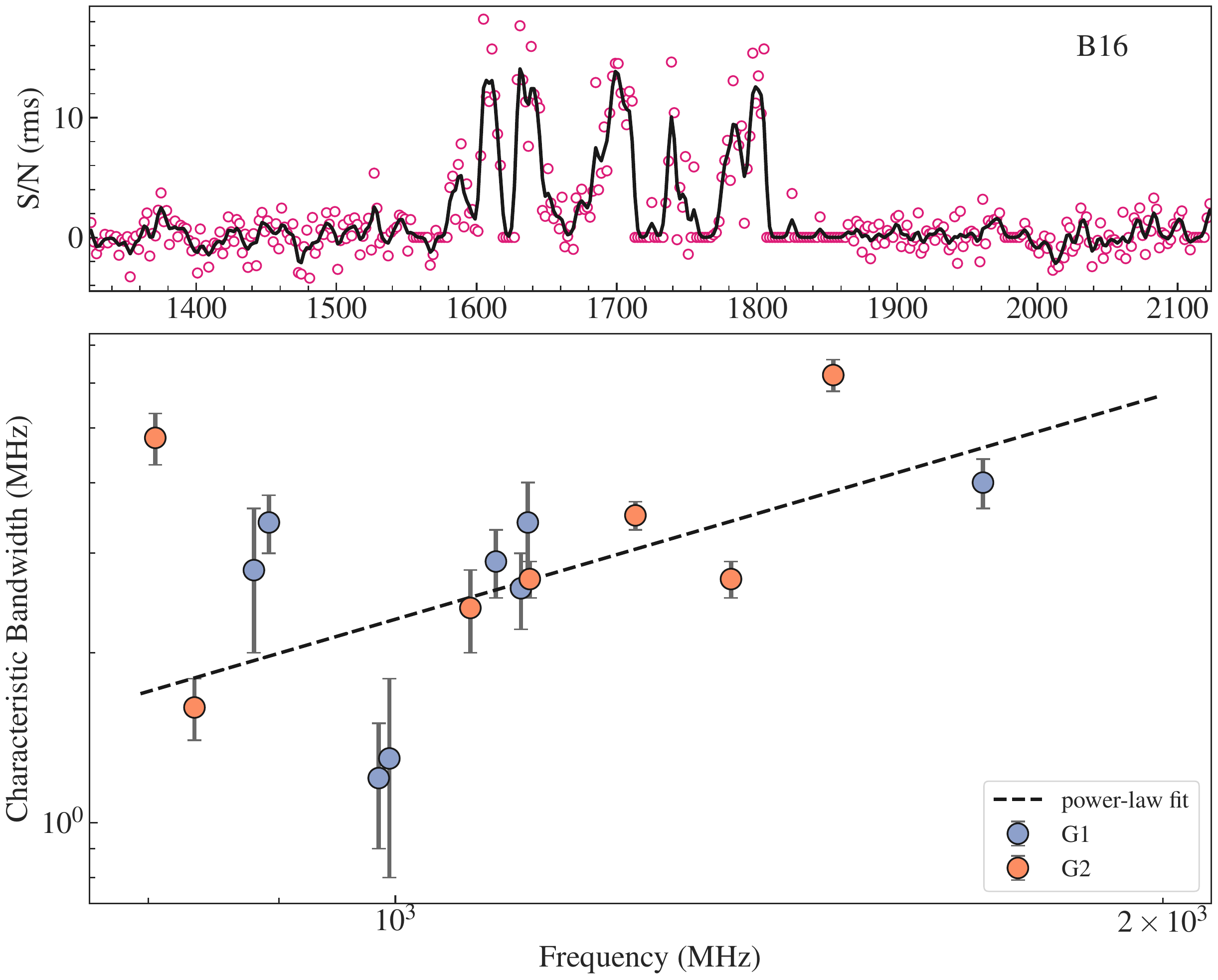} 
\caption{\emph{Top panel}: Spectrum of burst B16 showing abrupt edges (frequency resolution = 2~MHz). The solid line represents the Gaussian kernel density estimate of the spectrum. \emph{Bottom panel}: Characteristic bandwidth (BW$_{\rm sc}$) of the burst spectra for FRB\,20180301A Parkes detections plotted against the central frequency of their spectral envelope. The best power-law fit BW$_{\rm sc} \propto \nu^{1.3}$ is shown as the dashed line. }
\label{fig:B16_ACF}
\end{figure}

All Parkes-detected repeat bursts from the FRB\,20180301A source are band-limited, which makes determining their spectral extent challenging and, consequently, the burst spectrum. In addition, the burst spectrum of several bursts is largely non-contiguous because of the presence of RFI, which affects a large number of frequency channels. Moreover, the spectrum for many bursts has sharp edges (see Fig.~\ref{fig:B16_ACF}) , which prevents us from modelling it using any analytical function. Instead, we rely on the spectral extent reported in Table~\ref{tab:burst_properties} for each burst to ensure that we include all the burst power and maintain consistency across our analysis. Using a boxcar filter, we first determine the temporal on-pulse region of the burst profiles and select the best-matched template boundary as the region. We then designate the remaining time bins as the off-pulse region ($\sim$0.5\,s). To account for non-uniform instrumental frequency response, we normalise the data by subtracting the off-pulse mean and dividing it by the off-pulse RMS. We note that the dynamic spectra units have changed from absolute flux units to standard deviations of the off-pulse noise. The normalisation reduces the effects of channel outliers left over from RFI mitigation, which could otherwise affect the spectral baseline. Once normalised, the burst spectrum is generated by summing the on-pulse region in time for each frequency channel.

Although frequency-dependent brightness variations are observed in the spectra of several repeat bursts in our sample, these characteristics are unlikely to be due to radio signal scintillation along the line of sight. Both the NE2001 and YMW16 galactic models predict scintillation bandwidths on $\sim$kHz-scales at 1\,GHz along the line of sight of this FRB source \citep{Cordes:2002, Yao:2017}. These expected variations are much smaller than the frequency resolution (0.5--1\,MHz) of the Parkes-detected bursts. Nevertheless, we perform a standard ACF analysis to determine the ``characteristic bandwidth'' of the brightness variations in these bursts. The variation pattern in the spectral intensity changes with observing frequency. This change is characterized by the decorrelation bandwidth \citep{Cordes:1985}, typically defined as the half-width at half-maximum (HWHM) of the first lobe of the autocorrelation function of the spectrum \citep{Gwinn:1998, Ocker:2022}. 

We performed the ACF analysis on a sample of bursts with high S/N (>20) to avoid any bias introduced due to noise or RFI. We used the central frequency of the burst spectral extent to determine the evolution of the characteristic bandwidth (BW$_{\rm sc}$). The mean BW$_{\rm sc}$ was found to be 3\,MHz, with no clear trend of an increasing BW$_{\rm sc}$ with frequency. The obtained measurements were consistent throughout the monitoring period, and there was no discernible difference between the bursts in groups G1 and G2. However, we attempted to model the BW$_{\rm sc}$ evolution with frequency using a power-law function BW$_{\rm sc} \propto \nu^{\gamma}$ and obtained an index of $\gamma=1.3\pm0.5$, which is inconsistent with scattering arising in a turbulent plasma, where we would expect $\gamma \approx 4$. We conclude that the observed brightness variations were not due to interstellar scintillation and is inconsistent with multipath propagation. It is possible that the ACF method may not be reliable for determining BW$_{\rm sc}$ close to the frequency channel resolution. Figure~\ref{fig:B16_ACF} shows the obtained bandwidths along with the best-fit power law.

\begin{figure}
\centering
\includegraphics[width=\columnwidth]{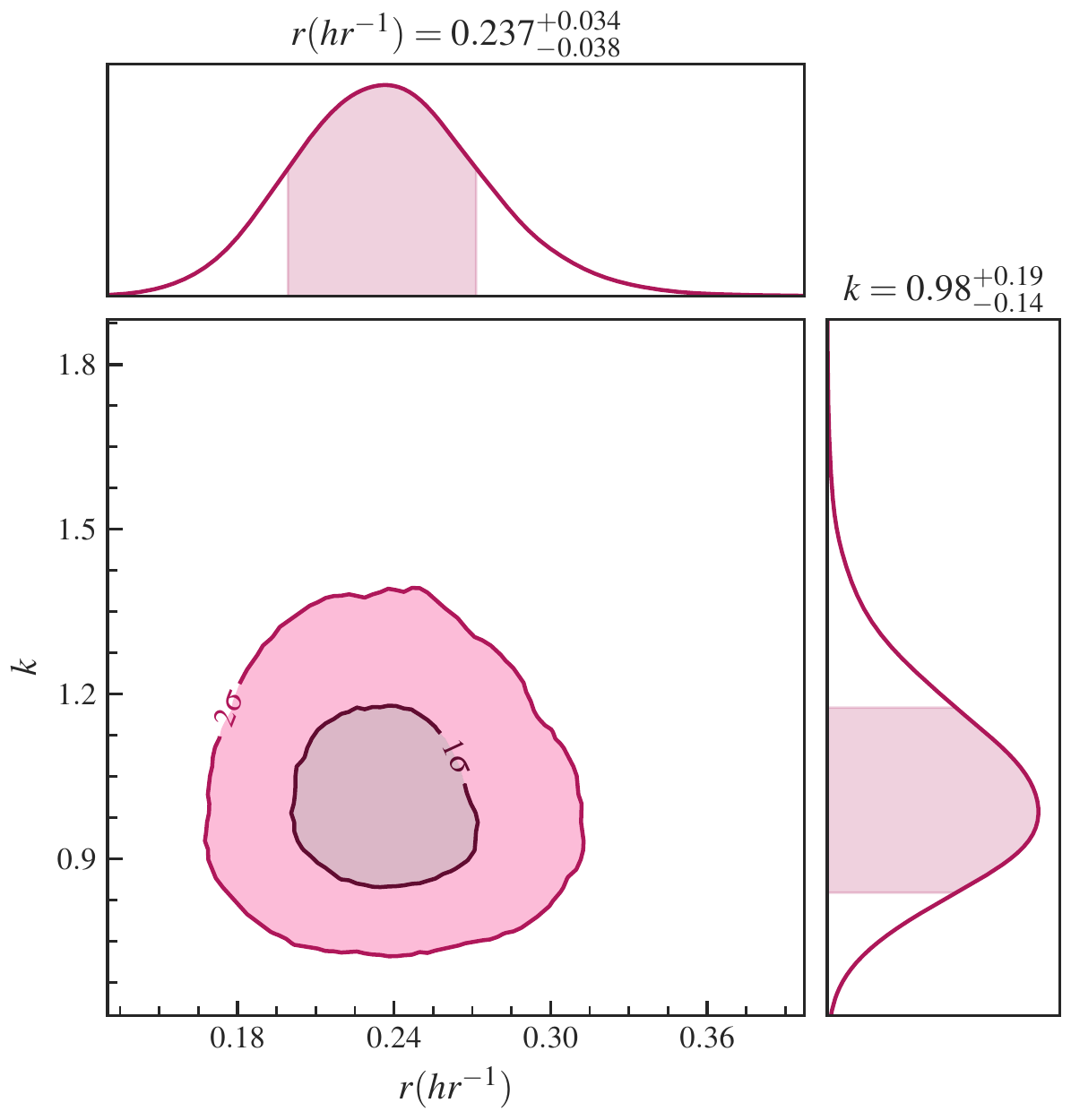}
\caption{Posterior distributions for burst rate $r$ ($\mathrm{hr}^{-1}$) and the shape parameter $k$ of the Weibull distribution using Parkes observations for the repeating source FRB\,20180301A. The bottom left panel shows the distributions of these parameters, with the contours representing $1\sigma$ and $2\sigma$ confidence levels for 2D Gaussians. The top and bottom right panels show the marginalised distributions for $r$ and $k$, respectively.}
\label{fig:rate}
\end{figure}

\subsection{Periodicity and burst rate}
Periodic modulation of burst activity is an important characteristics of FRBs that can provide direct insights into the underlying astrophysical processes responsible for their emission. In the case of the FRB\,20180916B source, studies using discrete Fourier transform searches on CHIME observations found a 16.33-d periodicity for its repeat bursts \citep{CHIME:2020_periodicity, Pleunis:2021_LOFAR}. We used a similar technique\footnote{\url{https://github.com/evanocathain/Useful_RRAT_stuff}} to search for both long-term and short-term periods in the FRB\,20180301A source activity. This technique has been successfully used to identify periodicity in rotating radio transients 
\citep[RRATs;][]{Morello+20MN}, robustly addressing the non detections. Specifically, we searched for periods in the ranges of 1 to 100 days and 1 ms to 100 s. However, our analysis did not reveal any significant harmonics, indicating no evidence for periodicity in the activity of FRB\,20180301A source over the Parkes UWL observing span. 

Repeating FRBs often exhibit periods of non-detection during long-term monitoring, which may be the result of unknown non-Poissonian process \citep{Connor+16MN}. To account for this and quantify the burst rates, we model the detected events of FRB\,20180301A in our Parkes observations using a Weibull distribution \citep{Oppermann+18MN}. The distribution function for event time intervals is given as
\begin{equation}
    \mathcal{W}(\delta|k,r)=k\delta^{-1}\left[\delta r \Gamma(1+1/k)\right] e^{-\left[\delta r \Gamma(1+1/k)\right]^k},
\end{equation}
where $\delta$ is the waiting times between adjacent events, $r$ is the burst rate, $\Gamma(x)$ is the gamma function and $k$ is its shape parameter. We use the package \software{BayesWeib\footnote{\url{https://github.com/RuiLuoAstro/BayesWeib}}} to sample the posterior distribution using a nested sampling algorithm. We assume log-uniform priors on the model parameters with $10^{-3} < r (\mathrm{hr}^{-1}) < 10$ and $10^{-2}<k<10$. We model the burst detections and the Parkes observation span (shown in Fig.~\ref{fig:timeline}), and obtain the mean burst rate $r=0.24\pm0.04\,\mathrm{hr}^{-1}$ and the shape parameter of the Weibull distribution $k=1.0\pm0.2$ with 95 per cent confidence levels. The posterior distributions of the model parameters are shown in Fig.~\ref{fig:rate}. Our analysis found that the burst rate of Parkes UWL observations is roughly five times lower than that of FAST observations i.e., $r=1.2^{+0.8}_{-0.7}\,\mathrm{hr}^{-1}$ \citep{Luo:2020}. The shape parameter $k$ is consistent with unity, which means that the Weibull distribution is identical to a Poissonian distribution. 

\section{Discussion}\label{sec:discussion}
\subsection{General burst properties}
One of the fundamental properties used for comparing competing emission mechanisms and FRB progenitor models is burst duration. Most FRBs exhibit a temporal envelope distributed within the range of $\approx$1--10\,ms \citep{CHIME:2021}. However, several bursts with a pulse envelope spanning $\sim\upmu$s scales have been detected from the FRB\,20121102A source, thereby extending the FRB temporal extents across several orders of magnitude \citep{Michilli:2018}. High-time resolution studies of many FRB sources have revealed bursts with temporal structures or sub-bursts down to a timescale of a few microseconds within the burst envelope \citep{Farah:2018, Cho:2020, Day:2020}. Burst structures with durations as short as $\lesssim$100\,ns have been discovered in the FRB\,20200120E source, which is associated with a globular cluster in the M81 spiral galaxy \citep{Majid:2021, Kirsten:2022}. These nano-shots are morphologically similar to the ones emitted by the Crab pulsar, suggesting an observational link between FRBs and radio pulsar giant pulses \citep{Nimmo:2021, Nimmo:2022}. Pulse structures on such short timescales allow for critical evaluation of emission models, as it imposes stringent constraints on the size of the FRB emitting region. Such similarity indicates a magnetospheric origin for the burst emission. Alternatively, if the emission originates from relativistic shocks located far away from the progenitor, the similarity of the pulse structures on short timescales suggests that a small frontal region dominate the emission process \citep{Metzger:2019}.

Bursts from repeating FRB sources exhibit longer temporal extents and more complex morphologies in comparison to those from apparently non-repeating sources \citep{Fonseca:2020, Pleunis:2021}. In the most extensive studies of repeating FRBs, specifically FRB\,20121102A and 20201124A, conducted with FAST, the width distribution of repeat bursts has been found to follow a log-normal distribution on ms-scales, centering around 4--7\,ms \citep{Li:2021, Zhou:2022}. However, the burst width distribution on $\upmu$s--scale has not yet been thoroughly investigated. Bursts B10 and B16 exhibit extremely narrow temporal widths of $\lesssim100\,\upmu$s and stand out as outliers within the overall Parkes-detected sample of FRB\,20180301A repetitions. The limited temporal resolution of our observations prevents probing the temporal burst components in greater detail. Similar sub-population of $\upmu$s--scale bursts has also been discovered in other repeating FRBs \citep{Michilli:2018, Nimmo:2021}, implying a resemblance in their physical origins. While the complete distribution of burst widths is unknown, many bursts may be missing due to inefficient search methods at these time scales. A more comprehensive coverage of pulse widths, spanning across $\upmu$s--ms scales at high-time resolution, will allow us to discover any bimodality in the width distribution and whether there is a distinct mechanism responsible for the bursts with temporal scale $\lesssim100\,\upmu$s.

\begin{figure}
\centering
\includegraphics[width=\columnwidth]{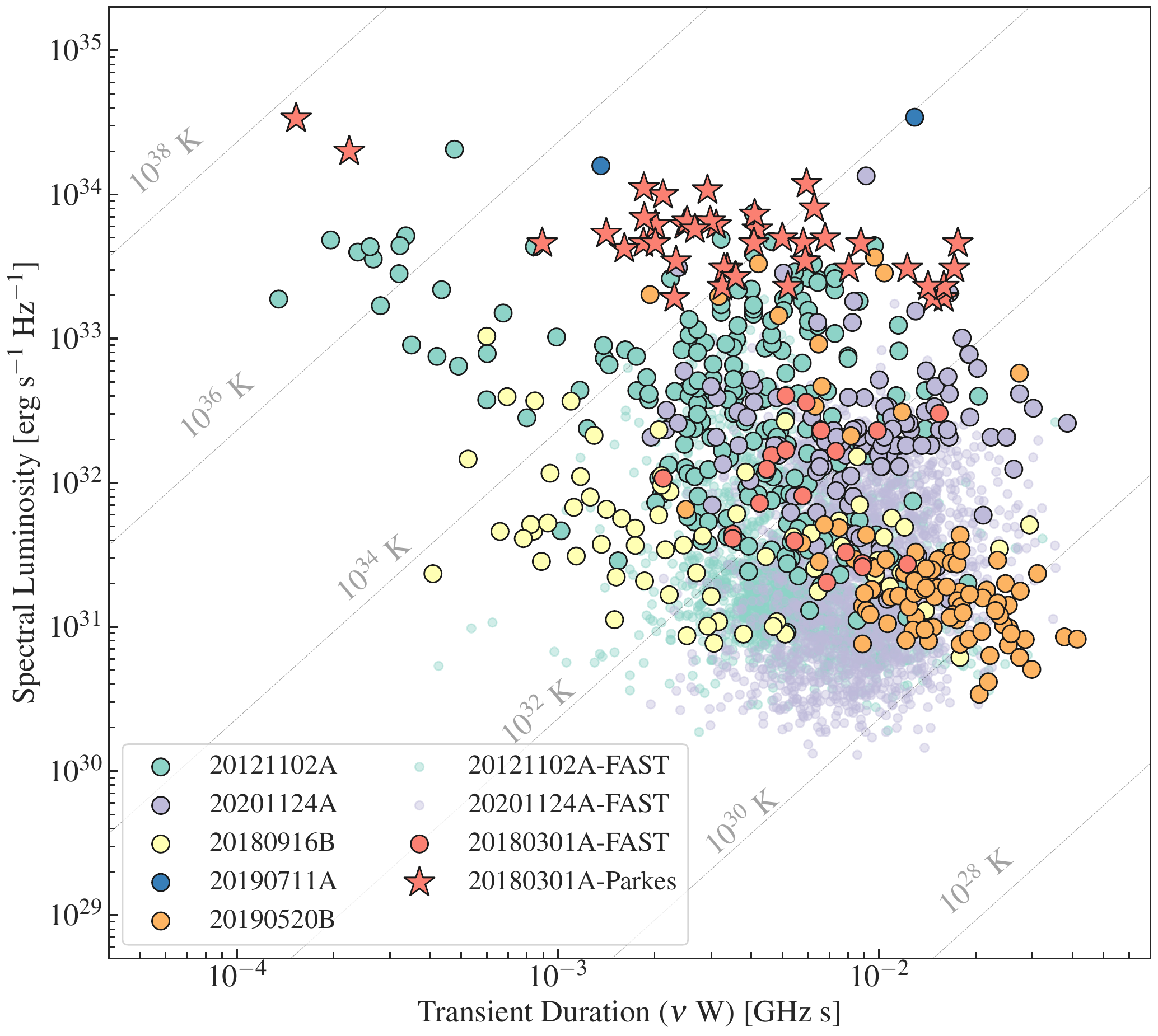} 
\caption{Repeating FRB phase space: The inferred isotropic peak spectral luminosity of repeat bursts and its relationship to the burst duration (product of central emission frequency and pulse width). The black dashed lines indicate the brightness temperature. Only published localized FRBs with known redshifts are used for this analysis \citep{Luo:2020, Hilmarsson:2021_CP, Li:2021, Kumar:2022, Laha:2022, Lanman:2022, Marthi:2022, Nimmo:2022, Niu:2022, Xu:2022, Thomas:2022}. Star markers indicate Parkes-detected repeat bursts from the FRB\,20180301A source, as reported in this paper.
}
\label{fig:tps}
\end{figure}

The exact definition of what constitutes an FRB has yet to be determined. Many of the observational properties of these bursts span several orders of magnitude, with much of the parameter space yet to be fully explored \citep{Petroff:2022}. One burst property that has been proposed as the definition of a ``classical'' FRB is the brightness temperature, $T_{\mathrm{b}}$ \citep{Zhang:2022}. It has been suggested that a critical brightness temperature could serve as a discriminatory parameter between typical and atypical bursts \citep{Xiao:2022}. Here, we report repeat bursts B10 and B16 with the highest brightness temperature detected among all repeating FRBs to date. Both bursts exhibit $T_{\mathrm{b}} \sim10^{37}$--$10^{38}$\,K, which further pushes the boundaries of the ``extreme scenario'' required by coherent radio emission mechanisms. Such high values of $T_{\mathrm{b}}$ is challenging to explain within the framework of the proposed theoretical models \citep{Xiao:2022, Zhang:2022} and there have been attempts to relate it with pulsar radio emission, often invoking reconnection-driven models \citep{Lyutikov:2021}. In Fig.~\ref{fig:tps}, we present the phase space diagram of repeating FRBs, comparing Parkes-detected FRB\,20180301A bursts with the whole population. For this comparison, we confined our analysis to published repeating FRBs with known redshifts. Notably, we excluded the bursts from FRB\,20200120E due to their significantly reduced luminosity by three orders of magnitude, attributed to the proximity of the source. We considered the entire burst envelope of the burst as the definition of the pulse width. It is worth noting that the brightness temperature associated with sub-bursts from FRB sources can attain even higher values, reaching $T_{\mathrm{b}}$ $\sim10^{40}$\,K \citep{Nimmo:2021}.  

One of the characteristic morphological features of repeating FRBs is frequency drifting of sub-pulses \citep{Gajjar:2018, Hessels:2019}. In nearly all cases, this drifting occurs from higher to lower frequencies, though there are instances of bursts with ambiguous morphologies \citep{Marthi:2020, Pleunis:2021_LOFAR}. Several bursts in the Parkes sample exhibit downward-drifting sub-pulses, displaying a significant variability in drift rates that aligns with trends seen in other repeating FRBs \citep{Platts:2021}. The wide-band coverage facilitated by the UWL system provides a unique advantage in measuring drift rate evolution. Unlike previous studies that relied upon drift rate estimates derived from multiple radio telescopes at different frequencies, potentially introducing instrumental biases, the UWL offers the unprecedented capability to conduct drift rate evolution measurements using a single instrument. 

The inferred drift rates $|d\nu /dt|$ in the Parkes sample are found to increase with observing frequency, similar to the linear evolution observed for FRB\,20121102A and FRB\,20180916B \citep{Marazuela:2021, Platts:2021, Sand:2022}. A power-law model to describe the drift rate evolution within the Parkes sample is in agreement with the best-fitting model $d\nu /dt \propto \nu^{2.48}$ obtained for bursts from other repeating FRBs \citep{Zhou:2022}. This consistent behaviour of sub-pulse drifting, exhibiting uniform sign and frequency dependence across the broader FRB population, strongly indicates the possibility of a shared underlying mechanism \citep{Metzger:2022}. The phenomenon of sub-pulse drifting in FRBs is usually interpreted in terms of radius-to-frequency mapping, where the emission originates within the magnetosphere of the progenitor source \citep{Lyutikov:2020}. Other models include the triggered relativistic dynamical model, which assumes a narrow-band emission process intertwined with up-to-relativistic motions within the source \citep{Rajabi:2020, Chamma:2021}.

Repeating FRBs display a wide range of burst rates, exhibiting variations in both frequency and energy dependence \citep{CHIME:2023_repeaters}. While some sources, like FRB\,20121102A and 20180916B, exhibit periodic modulation in their burst activity, thus confirming long-term stability, \citep{CHIME:2020_periodicity, Rajwade:2020}, many others are prolific burst emitters without exhibiting any clear signs of periodicity \citep{Nimmo:2023}. In contrast, the active source FRB\,202021124A displayed an unusual level of burst activity that abruptly ceased for an extended period \citep{Xu:2022, Zhou:2022}. This complicates our understanding of the stability profiles of the repeating FRB sources and the underlying physical mechanisms driving these bursts. Most sources have low repetition rates, requiring extensive follow-up observations with sensitive telescopes to successfully detect bursts \citep{Kumar:2019}. Our extensive monitoring of the FRB\,20180301A source spanning two-and-a-half years has revealed a fairly consistent burst rate, devoid of any evidence of periodic modulation. The inferred burst rate follows a Poisson distribution similar to other repeating sources, indicating a common underlying mechanism for burst production \citep{Cruces:2021}. Parkes monitoring suggests that the activity level and emission patterns of some FRB sources might exhibit a greater degree of constancy compared to others. The absence of periodicity, on the other hand, emphasises the need for more sensitive and long-term monitoring.

Repeating FRBs also display a strong correlation between their emission activity and the observing frequency \citep{Aggarwal:2020RNAAS}. The wide-band coverage of the UWL also makes it a powerful tool for studying the spectral dependence of burst emission across a bandwidth of 3.3\,GHz. This eliminates the need for simultaneous observations with multiple radio telescopes, reducing potential instrumental biases \citep{Law:2017}. Our monitoring of the FRB\,20180301A source revealed a lack of detectable radio emissions above 1.82\,GHz. The majority of bursts were centred around $\sim$1.1\,GHz, suggesting a preferred emission frequency for our observation epochs. However, this is not necessarily the case for other FRB sources. For example, FRB\,20190520B regularly emits bursts at frequencies spanning 3-7\,GHz \citep{Thomas:2022}. Other repeating FRBs also display burst emissions that span a range of radio frequencies \citep{Gajjar:2018}. It is possible that in the case of FRB\,20180301A, radio emission at higher frequencies is not sufficiently bright, and further sensitive observations are needed to thoroughly investigate the presence of any missing high-frequency emission.

\subsection{Polarization variability}
FRB\,20180301A is the first known repeating FRB source to display substantial variations in the polarization PA across its pulse profiles. The observed PA swings in the FAST-detected repeat bursts closely resembled those seen from radio pulsars and magnetars, implying that these bursts are likely of magnetospheric origin \citep{Luo:2020}. We found no evidence of a significant PA variation within the Parkes-detected repeat bursts. All polarized UWL bursts, including some relatively bright ones, displayed nearly constant polarization PA across the burst envelope. It might be that the PA variations are not a prevailing emission characteristic of FRB\,20180301A source, which is consistent with the diverse polarization properties observed in other repeating FRBs. In recent observations of the prolific repeating source, FRB\,20201124A, the majority of bursts showed no PA variation \citep{Hilmarsson:2021_CP}; however, only a select few displayed substantial variations, ranging up to $\sim$50$\degr$ \citep{Kumar:2022, Xu:2022}. It is also possible that the electromagnetic radiation traverse through multiple pathways within the magnetosphere of the FRB progenitor resulting in such unusual PA variations in a fraction of repeat bursts.

Circular polarization (CP) is one of the remarkable characteristics observed in some of the bursts from repeating FRBs. Detection of CP in radio emission, in addition to non-varying PA, was previously thought to be a distinctive marker of one-off FRBs \citep{Chatterjee:2021}. Extensive monitoring of many active repeating FRBs (20201124A, 20121102A and 20190520B) in the last few years has revealed CP with a wide spread of fractional strengths ranging up to 70 per cent \citep{Hilmarsson:2021_CP, Kumar:2022, Xu:2022, Jiang:2022, Feng:2022_circular}. However, CP is not a dominant feature in majority of bursts from these sources, implying that a completely different emission mechanism could be responsible for its generation or conversion from the linearly polarized component in the propagating medium \citep{Gruzinov:2019, Vedantham:2019, Kumar:2022_gfr}. For the original FRB\,20180301A detection with the Parkes MB, \cite{Price:2019} reported significant CP in the burst spectra. However, it was not clear from the initial measurement whether the CP component in the burst spectrum was due to instrumental leakage or was intrinsically associated with the FRB signal. In the subsequent follow-up with FAST, \cite{Luo:2020} found no presence of CP in their sample of repeat bursts. We also do not detect a significant (> 15 per cent) CP fraction in any of the UWL-detected bursts from FRB\,20180301A. This is consistent with the general consensus that, while CP may be an essential feature of repeating FRB emission, it is not common, thus implying a changing progenitor environment.

We find significant temporal variations in the Faraday RM of the Parkes-detected repeat bursts with a measurable polarization. There is tentative evidence of linear variations over the course of our follow-up period. Linear regression indicates a monotonic increase in the RM during 2020--2022 with a variation rate of $98\pm55$\RMtrendunits. \cite{Luo:2020} also reported a potential linear trend with a deviation rate of $\sim$21\RMtrendunits~in the FAST-detected bursts on a timescale of a few months. Nonetheless, we concur that the evidence of such a linear variation is not concrete based on our small sample size. Linear regression cannot adequately model the RM variations in some notable repeating FRB sources 20190520B and 20201124A \citep{Xu:2022, Thomas:2022}. The linear variation rate also does not explain the absence of LP in the majority of Parkes/UWL bursts. An intra-channel depolarization is unlikely because it would imply that the RM was fluctuating with $> 10^4$\RMunits during the monitoring period. Most likely, those bursts are either intrinsically depolarized or are being affected by propagation effects, which may have resulted in a lower LP signal strength \citep{Feng:2022}. 

Faraday RM variations in repeating FRBs are a recurring characteristic that has been observed on a wide range of timescales. FRB\,20121102A, which holds the record for the highest detected RM in any FRB source at $\sim10^5$\RMunits, exhibited a non-linear decline at frequencies > 1.5\,GHz \citep{Michilli:2018}. The RM of this FRB source has been consistently decreasing at an average rate of 12 per cent per year \citep{Hilmarsson:2021, Plavin:2022, Feng:2023_ATel}. The other well-studied repeating source, FRB\,20180916B, displayed RM evolution in particular periods. Following an extended phase marked by stochastic variations, this source displayed a secular increase in RM with a linear variation rate of $72\pm2$\RMtrendunits \citep{Mckinven:2022}. Similarly, in one of the CHIME-discovered sources, FRB\,20190303A, the RM has changed by $\sim500$\RMunits over a span of two years, displaying intervals of secular evolution \citep{Feng:2022, Mckinven:2023}. Another such source, FRB\,20181119A displayed substantial RM variability of $\sim860$\RMunits over a period of six months \citep{Mckinven:2023}. An irregular yet significant temporal variation in Faraday RM was discovered in FRB\,20201124A with $\Delta \textrm{RM}\sim500$\RMunits within a span of a few months. Curiously, the RM variation also ceased suddenly in a peculiar manner, characterized by a remarkably constant RM in the days preceding the cessation of burst emissions \citep{Xu:2022}. Lastly, extreme RM variations of $\pm300$\,rad\,m$^{-2}$\,day$^{-1}$ on week-long timescales has recently been observed in FRB\,20190520B \citep{Thomas:2022}. It is also possible for Faraday RM variation to occur as a function of the observing frequency if the burst radiation traverses different regions within the progenitor's magnetosphere \citep{Plavin:2022}. Nonetheless, such effects might be less discernible and challenging to probe, given the already significant temporal RM variations exhibited by FRBs. It is becoming evident within the context of repeating FRBs that there are possibly different regimes or time intervals in which regular or irregular variation in Faraday RM can be induced \citep{Mckinven:2022}.

Furthermore, the Faraday RM variation in FRB\,20180301A source indicates an unambiguous sign change during 2019--2020 (see Fig.~\ref{fig:dm_rm_evolution}). It is unclear whether the decline from the FAST-detected average RM $=550$\RMunits, to the Parkes-detected RMs was linear or abrupt, owing to the absence of detectable polarized flux in the initial bursts B01-B09. The evident RM sign-reversal also lends support to the authenticity of the RM\,$= -3163$\RMunits reported for the original FRB\,20180301A burst \citep{Price:2018}. In addition, this apparent RM indicates another sign-reversal between the 2018 Parkes detection and the 2019 FAST detections. A particularly intriguing aspect is the long-term ascending RM trend apparent in the Parkes-detected repeat bursts, hinting at the potential occurrence of a future sign reversal in the ensuing months. The phenomenon of sign reversals in Faraday RM has also been confirmed in recent observations of another repeating FRB source 20190520B. The observed RM for this source changed from $1.3\times10^4$\RMunits to $-2.4\times10^4$\RMunits within a span of five months \citep{Thomas:2022}. These occurrences of sign reversals are likely indicative of changes in the orientation of the parallel component of the magnetic field along the line of sight. 

Of the bursts in our sample, 70 per cent exhibit no detectable polarized emission, compared to 36 per cent in the FAST sample \citep{Luo:2020}. Notably, all high-frequency ($\gtrsim$ 1.2\, GHz) bursts in our sample are polarized, while most low-frequency bursts are not (see Figure\,\ref{fig:timeline}). Two exceptions, B08 and B27, are low-significance events with S/N $\lesssim$8, making reliable measurement of polarization infeasible. This suggests depolarization as a function of observing frequency, supported by a trend of decreasing degree of LP with burst emission frequency in our sample of polarized bursts. \citet{Feng:2022} showed that such a trend is common among several repeating FRBs (including FRB\,20180301A) based on available multi-frequency observations. This phenomenon is explained by an RM scatter model caused by multipath propagation in the inhomogeneous magneto-ionic environment. We obtained a similar RM scatter parameter $\sigma_{\rm RM}$ for the Parkes-detected repeat bursts, although with considerable measurement scatter along the modelled trend. This might suggest that like Faraday RM, the $\sigma_{\rm RM}$ might also experience temporal variations or the RM scattering parameterization is not adequate.

\subsection{DM variability}
Variations in the observed DM are common among repeating FRB sources due to the complex morphology of burst signals. However, distinguishing the higher-order non-dispersive effects from genuine DM variations caused by changes in electron density can be challenging. In addition, instances of short-term ($\sim$day) stochastic DM variations have also been observed in several repeating FRBs, further complicating the dispersion analysis \citep{Xu:2022}. To address these challenges, novel DM estimation techniques that optimise the burst structure have been developed \citep{Seymour:2019, Lin:2022}. These methods minimize the non-dispersive effects, allowing for more robust measurement of the signal dispersion and variations caused by changes in electron density.

The primary evidence for DM variation in repeating FRBs comes from the extensively studied source of FRB\,20121102A. Since its initial discovery, the FRB source has exhibited a secular upward trend in the average measured DM, amounting to $\sim$1\DMtrendunits \citep{Spitler:2016, Hessels:2019, Oostrum:2020, Li:2021}. However, recent observations of the FRB source, following a gap of three years, have revealed a reduction in the DM by $\sim$10\DMunits \citep{Wang:2022ATel, Li:2021}. While there has been ample evidence of DM variations on a relatively short timescale (<1 day), no pronounced secular variations extending over the span of months to years have been found in the similar prolific repeater sources FRB\,20190520B and FRB\,20201124A \citep{Thomas:2022}. In the case of FRB\,20180916B, which also shows periodic modulation in the burst activity, no significant DM variation has been observed so far. For this source, stochastic DM variations across bursts are constrained to a limit of $\leq 1$\DMunits \citep{Sand:2022, Mckinven:2022}. Similarly, for the M81 repeater FRB 20200120E, DM variations are constrained to be $\leq 0.15$\DMunits \citep{Nimmo:2023}. Assuming that all FRBs have a similar mechanism governing DM variability, the differences observed across the FRB population suggest that most FRB progenitors are not surrounded by dense supernova remnants or, if they were at some point, the remnant has faded enough to generate any DM variation. This also suggests that any residual magnetar wind does not affect the propagation of burst signals \citep{Marazuela:2021}. These observations disfavour binary wind models and support a cleaner local environment compared to the extreme magneto-ionic environments of FRB\,20121102A \citep{Nimmo:2023}.

Our monitoring of FRB\,20180301A source has revealed compelling evidence of secular evolution within its observed DM. Unlike FRB\,20121102A, the DM of this FRB source exhibits a decreasing trend over the course of our monitoring period, indicating of a reduction in column density during this time-frame. Notably, the magnitude of DM variation evident within this FRB is relatively larger than that of FRB\,20121102A \citep{Li:2021}. However, it is challenging to pin down the beginning of this evolutionary trend. 
The inferred trend might also be influenced by systematic factors stemming from the spread of DMs of the repeat bursts during a relatively active phase, e.g., the bursts detected during 2021 July--Sep. The average DM for G1 bursts is in agreement and within 2$\sigma$ of the average DM$_\mathrm{struct}$ measured for FAST-detected bursts from this FRB source in 2019 \citep{Luo:2020}. This intriguing similarity suggests that the observed DM trend commenced alongside the G1 bursts.

The observed DM variations across the population of repeating FRBs are not consistent. It is possible that these variations materialise during different phases of the progenitors' activity cycle, potentially associated with changes in their local environment. Not all repeating FRBs exhibit similar levels of burst activity, and many of these sources have not been extensively studied across the radio band \citep{CHIME:2023_repeaters}. Similar to the observed Faraday RM, the DM could exhibit stochastic variations during some phases while showcasing secular variations during others. Furthermore, the existence of periodic modulation correlated with the DM variation can help to discriminate models that involve binary systems. The potential linkage between periodic oscillations and DM variation can be decisively investigated through comprehensive, long-term monitoring with sensitive telescopes.

\subsection{DM-RM correlation?}
The observed DM of FRBs serves as a measure of the free electron density along the line of sight. Additionally, the detection of linear polarization facilitates the measurement of Faraday RM, enabling an inference of the the electron-density weighted average of the parallel component of the magnetic field strength. While both of these measures encompass contributions originating from various plasma components along the signal path (e.g., interstellar medium, intergalactic medium, host galaxy, and local plasma), significant variations can primarily be attributed to the local plasma component. This is grounded in the improbability of the other contributing components undergoing significant evolution within the time scale of our observations \citep{Yang:2022_RM}. Consequently, a time-evolving DM and RM can provide valuable insights into the immediate environments of FRB progenitors, and the underlying mechanisms that drive their central engine \citep{Mckinven:2022}. 

The Galactic DM contribution along the direction of FRB\,20180301A, according to the NE2001 and YMW16 models, is 152 and 254\DMunits respectively \citep{Cordes:2002, Yao:2017, Price:2021}. Additionally, the Milky Way Halo is expected to contribute $\sim$39\DMunits \citep{Prochaska:2019_haloes}. By taking the average of the both model estimates, denoted as $\rm{DM}_{\rm{ISM}}$, this translates to an excess DM contribution, $\rm{DM}_{\rm{excess}} \sim $275\DMunits. Likewise, we estimate the Galactic RM contribution along the direction of the FRB to be $\rm{RM}_{\rm{MW}} = 44\pm19$\RMunits \citep{Hutschenreuter:2022}. There is also a small contribution stemming from the Earth's ionosphere impacting the observed RM. However, it is not significant for our case, and during our observations, we find that $\rm{RM}_{\rm{iono}} \lesssim 1$\RMunits \citep{Sobey:2019}.

We have found preliminary indications of a correlation between the DM variation and the change in RM of the FRB\,20180301A source in our Parkes observations. The correlated variability implies that the DM-RM evolution might be dominated by fluctuations in the free electron density $n_{e}$, rather than variations in the parallel component of magnetic field $\rm{B}_{\parallel}$ of the local environment. If this interpretation holds, the DM-RM correlation provides a direct avenue for inferring the magnetic field of the Faraday-active medium, given as
\begin{equation}\label{eqn:b_par}
\langle B_\parallel^{\rm local} \rangle = 1.23 \left( \frac{{d\rm RM/dt}}{{d\rm DM/dt}}\right)(1+z) = 60\pm34\,{\rm \upmu G}
\end{equation}
The inferred value is consistent with a wide range of estimates ($B_\parallel^{\rm local}$ up to several hundred $\rm \upmu G$) reported for other repeating FRBs and pulsars \citep{Hamilton:1985, Rankin:1988, Johnston:2005, Mckinven:2023}. The requirement that $n_{e}$ is positive implies that the RM associated with the local medium, $\rm{RM}_{\rm{local}}$ can not change sign as a result of the secular evolution in $n_{e}$. Assuming that the observed RM is predominantly contributed by $\rm{RM}_{\rm{local}}$, as observed in other FRB sources, it suggests that the local DM contribution, $\rm{DM}_{\rm{local}}$ is diminishing during the current evolutionary phase and is expected to approach $\rm{DM}_{\rm{local}}\sim$0 in upcoming months. An alternate scenario could involve concurrent changes in $B_\parallel^{\rm local}$ along with the local $n_{e}$. Further comprehensive monitoring will help in substantiating the observed DM-RM trend and addressing the degeneracy between the parameters characterizing the local environment.

\subsection{Possible astrophysical scenarios}
Many astrophysical scenarios have been proposed in the literature to explain the complex dispersion and polarization features observed in FRB signals \citep{Yang:2022_RM, Zhang:2022}. Here we focus on the models that are relevant to our observations of the FRB\,20180301A source. Given the spectral dependence of burst polarization properties, a scenario involving multi-path propagation through a turbulent and dense magneto-ionic plasma screen has been proposed to explain the signal depolarization and RM fluctuations \citep{Beniamini:2022}. The presence of an inhomogeneous screen can also lead to strong scintillation effects on the burst signal. This model has been used to demonstrate the observed effects in FRB\,20190520B \citep{Thomas:2022}. In this model, the effects observed are attributed to wave propagation in the local progenitor medium, which explains why such complex and extreme features manifest in only a few of the FRB sources, as these characteristics are not intrinsic to the FRB progenitor itself. 

Such a magneto-ionic plasma screen could have originated in several possible ways. A natural way to explain the observed secular DM and RM evolution is if the FRB source is surrounded by an expanding supernova remnant (SNR) or an ejecta from compact binary mergers \citep{Piro:2016, Margalit:2018, Zhao:2021_merger}. As the SNR ejecta evolves over time, the resulting forward and reverse shock wave interactions with the immediate environment result in a secular monotonic DM and RM evolution. A wide range of $d {\rm DM} /dt$ can exist depending on age of the ejecta as well as its evolutionary phase \citep{Piro:2018}. Given the observed decreasing DM, the models suggest that the SNR ejecta surrounding the FRB source is in the early blastwave stage, where it dominates the DM evolution \citep{Zhao:2021}. However, it is important to note, that such a correlated DM-RM evolution has not yet been observed in most repeating FRBs. This suggests that, on a first-order basis, not all FRBs necessarily conform to the SNR model. \citep{Metzger:2017, Yang:2017, Margalit:2018_metzger}. One plausible explanation is that the diversity within the age distribution of ejecta across different repeating FRBs may lead to varying evolutionary behaviours.

Other models suggest that the presence of a  high-mass companion with strong stellar outflows could be responsible for the observed Faraday RM variations in FRBs \citep{Zhao:2022}. For instance, a model proposed for FRB\,20201124A involves a binary system containing a magnetar and a Be star with a decretion disk \citep{Wang:2022}. The observed features show some analogy with the pulsed emissions from the Galactic binary system containing PSR B1259$-$63. The pulsar exhibit DM-RM variations and depolarization on short time scales as it approaches periastron in its binary orbit \citep{Johnston:1996}. A long-term trend of decreasing DM and absolute RM over two years is consistent with the FRB source being in a binary orbit with a massive object. It is possible that the Parkes observations coincide with a particular orbital phase, where DM and RM are decreasing, and the RM sign reversals may represent the periodicity of such an orbit. The DM could be decreasing over time due to the evolving positions of the objects within the system, thereby influencing the amount of matter traversed by the radio signal. In the case of PSR B1259$-$63, the Faraday RM reached a maximum value $\sim$10$^3$\RMunits and significantly reversed its sign around periastron. Another possibility is that the RM sign reversal is caused due to a dynamic and evolving magnetic environment around the FRB progenitor, for example due to its proximity to a massive black hole \citep{Thomas:2022}. While we have not observed an extremely large RM similar to FRBs 20121102A and 20190520B, this scenario can account for the RM sign reversal due to the change of magnetic field orientation in response to the progenitor's orbital motion around the black hole. A periodic pattern of such sign reversals could suggest that the progenitor is in a binary system, and such a pattern could be tested with future monitoring.

While the magneto-ionic plasma screen offers a plausible interpretation for the observed secular DM and RM evolution, it fails to account for other burst features. According to this model, depolarization arising from the scattering screen should result in a significant CP fraction in the burst signal \citep{Beniamini:2022}. Except for the original Parkes detection, where the origin of the observed CP remains ambiguous between astrophysical and instrumental factors \citep{Price:2018}, no significant CP fraction has been observed in bursts from the FRB\,20180301A source. Moreover, a magnetized scattering screen alone can not explain the observed PA swings in the FAST-detected bursts which have been proposed to be rooted in magnetospheric origins \citep{Luo:2020}. It is possible that multiple magnetized screens, each endowed with distinct characteristics, exist between the FRB progenitor and Earth, with different screens dominating the RM and DM evolution \citep{Beniamini:2022}.

\subsection{Potential PRS association}
The unusual Faraday RM variation and spectral depolarization observed in the FRB\,20180301A repeat bursts suggest the presence of a magneto-ionic plasma surrounding the FRB source, which could be associated with a co-located PRS. To date, only two other repeating FRBs, which are also prolific in their burst activity, have been associated with compact PRS sources. FRB\,20121102A was found to be co-located with a compact PRS (< 0.7\,pc in size) with a flux density of $\sim 180$\,$\upmu$\,Jy at 1.7\,GHz \citep{Chatterjee:2017, Marcote:2020}, whereas FRB\,20190520B is observed to be co-located with a PRS which is compact at VLA resolution (< 1.4\,Kpc) and have a flux density of $\sim 200$\,$\upmu$\,Jy at 3.0\,GHz \citep{Niu:2022}. Both of these FRB sources show similar burst properties to the bursts observed from FRB\,20180301A. Furthermore, they often exhibit extreme depolarization and RM scatter with $\sigma_{\rm RM} \gtrsim 30$\RMunits \citep{Feng:2022}. The measured $\sigma_{\rm RM} \sim7$\RMunits for the FRB\,20180301A bursts suggests a less magneto-ionic environment compared to FRBs 20121102A and 20190520B. 

No PRS was detected in the deep radio observations of FRB\,20180301A source with the VLA at 1.5 GHz, down to a $3\sigma$ flux density limit of $63$\,$\upmu$\,Jy\,beam$^{-1}$ \citep{Bhandari:2022}. However, the FRB source ($z=0.33$) is 85 and 44 per cent more distant than the FRBs\,20121102A ($z=0.2$) and 20190520B ($z=0.24$) respectively, meaning this non-detection limit only corresponds to excluding a source 1.2 times as bright as the FRB\,20121102A one. The VLA observations were thus not sensitive enough to detect a FRB\,20121102A-like PRS source at such high redshifts. Combined with the suggestion that any PRS may be fainter for FRB\,20180301A due to the less extreme environment, a non-detection therefore does not challenge models in which these sources all share a common progenitor type \citep{Law:2022}. The HG of FRB\,20180301A is not a dwarf galaxy like the other two hosts, indicating that it may be in a different environment. The observed similarities suggest that these FRB sources may be powered by young and active progenitors surrounded by dense and magnetized plasma.

On the other hand, CHIME-detected repeating sources that exhibit significant RM evolution and depolarization, such as that observed for FRB\,20180916B \citep{Mckinven:2022, Mckinven:2023}, do not appear to have any association with a compact PRS down to a $3\sigma$ flux density limit of $18$\,$\upmu$\,Jy\,beam$^{-1}$ with the VLA at 1.5 GHz \citep{Marcote:2020}. Rather, the low-frequency detection of the FRB source at 110\,MHz with the Low-Frequency Array (LOFAR) suggests a clear line-of-sight to the burst progenitor \citep{Pleunis:2021_LOFAR}. The observed RM variations from FRB\,20180916B indicate that significant fluctuations in the local magneto-ionic environment can still exist for sources that do not appear to be surrounded by highly magnetized plasma. Therefore, a larger FRB sample with localized HGs is required to investigate relationship, if any, between extreme polarization properties and the presence of a compact PRS. 

\begin{figure}
\begin{center}
\includegraphics[width=\columnwidth]{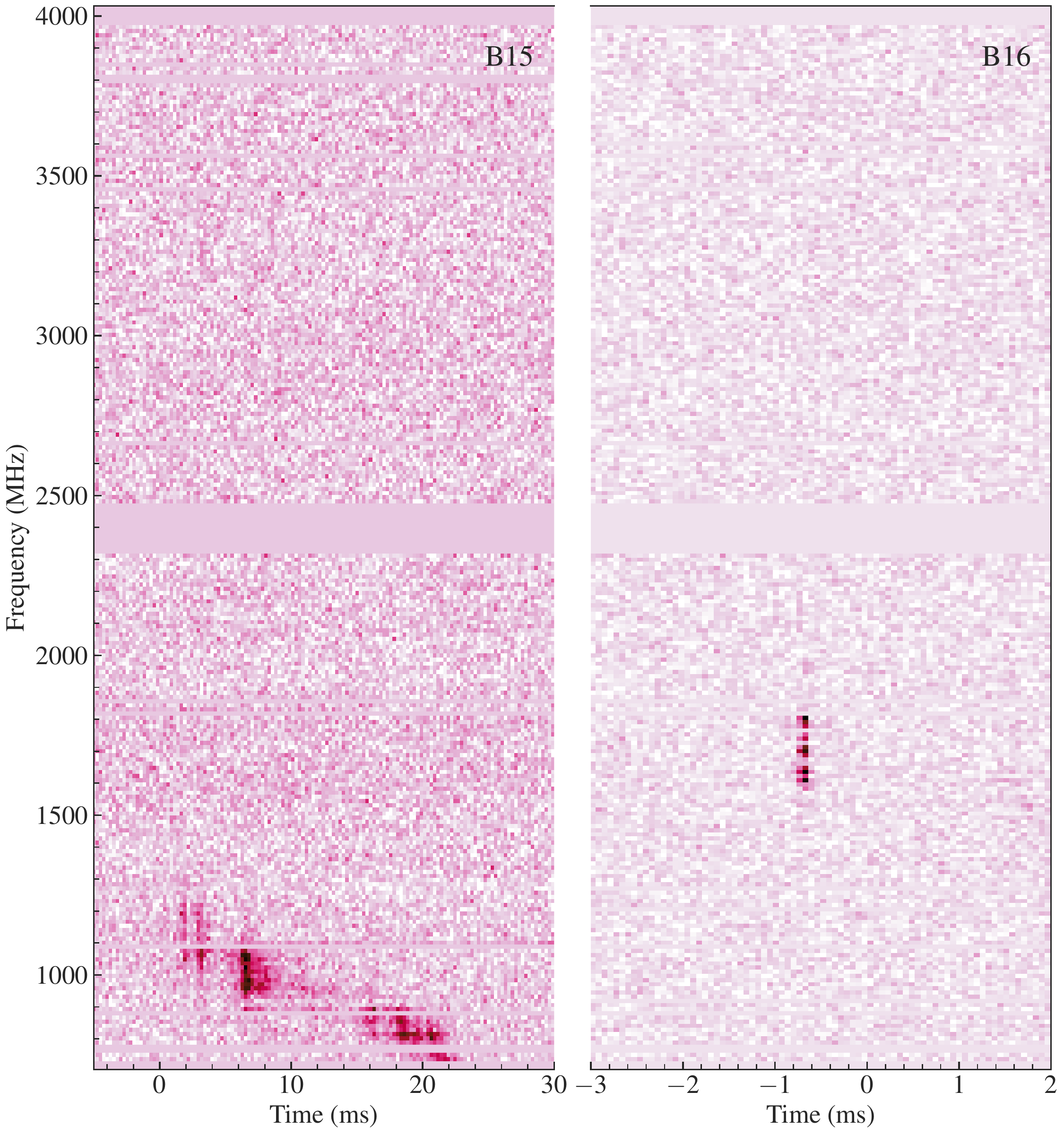}
\end{center}
\caption{Dynamic spectra of the Parkes-detected repeat bursts B15 and B16 from FRB\,20180301A source across the entire UWL band (frequency resolution = 13~MHz, time resolution = $256\,\upmu$s for B15 and $64\,\upmu$s for B16). Both bursts were detected in a single observing session with their time of arrival separated by 50\,min. The data have been dedispersed to the DM of 516.5 and 516.6\DMunits respectively. The dynamic spectra are normalized, and intensity values are saturated at the fifth percentile.}
\label{fig:b15_b16_contrast}
\end{figure}

\subsection{Dichotomy of repeat bursts?}
More contrasting features are now evident in detected burst signals from FRB sources. There is growing evidence of morphological differences among the FRB population, with bursts from repeating sources exhibiting larger temporal widths and narrower spectral extents \citep{Keane:2016, Pleunis:2021, CHIME:2023_repeaters}. However, it is possible that such studies are biased by instrumental limitations, resulting in the inference of a statistical dichotomy in the limited frequency bandwidth of the radio telescopes. Several distinction criteria have been proposed previously to distinguish repeating FRBs from the overall population. Examples of such criteria include a constant PA across the pulse profile, a high degree of LP and the absence of CP in the radio signal. However, extensive monitoring campaigns have revealed properties that challenge these criteria \citep{Luo:2020, Kumar:2022, Xu:2022}. While a broad classification based on polarization properties does not conclusively support a clear dichotomy between repeating FRB sources those which do not repeat, attributes such as a constant PA or the absence of CP tend to manifest more frequently in repeating FRBs. As such, these properties can still be indicative, particularly when considering ensemble of sources together. A comprehensive analysis of burst properties across different radio frequencies may provide more definitive evidence of differences among the FRB population.

The contrast in burst properties between the Parkes-detected repeat bursts B15 and B16 from the FRB\,20180301A source is striking. These two bursts, separated by $\sim$50\,mins with no other bursts detected during the entire 2.5-hr observing session, provide a clear example of the diversity within the repeating FRB population. The dynamic spectra of both UWL bursts are shown in Fig.~\ref{fig:b15_b16_contrast}. Both bursts are morphologically distinct. Burst B15 is a canonical repeat pulse with standard morphological features such as multiple components and downward sub-pulse drifting. In contrast, burst B16 has a remarkably narrow temporal envelope $\sim100\upmu$s with steep spectral cutoffs on both edges. Moreover, while no detectable polarization signal is present in burst B15, B16 exhibits 80 percent linear polarization. We also find that the subsequent bursts are polarized, prompting an intriguing question: did the emission mechanism switch or change after burst B15? These two bursts serve as prime examples illustrating how the emission from repeating FRBs can exhibit distinct spectral, temporal and polarimetric properties over a time scale of $\lesssim$1\,hr. 

The distinct frequency regimes of both bursts also raise the question of whether we might have missed other FRBs due to the narrow bandwidth of observing radio telescopes. This bias can significantly impact the calculation of true FRB detection rate. Since the spectral characteristics of FRBs can vary significantly from burst to burst, band-limited telescope observations may overlook bursts that have different spectral properties than those detected within the observing band. This could lead to a bias in the measured burst properties and a limited understanding of the physical mechanisms driving FRB emission. Our UWL observations highlight the importance of monitoring FRBs over long timescales and across broad range of frequencies to capture the full range of their variability.

Furthermore, our sample contains three bursts (B37--B39) that exhibit spectral extents in the range of 0.7--1.0\,GHz. This finding demonstrates that, like polarization discriminators, the spectral extent is also an imperfect indicator of the repeating nature of FRBs. Due to the limited bandwidth of radio telescopes, it is often impossible to determine whether a detected signal spanning the telescope's bandwidth should be classified as a narrow-band or broad-band pulse. Moreover, the FRB spectral ``bandedness'' is not a precisely defined parameter, and there is no consensus on a standardized definition \citep{Aggarwal:2021}. When calculating FRB energetics, it is often assumed (as for Galactic radio pulsars) that the spectral extent of the received signal should have $\Delta \nu \sim \nu$ \citep{Zhang:2022}. However, $\Delta \nu / \nu$ exhibit a wide range of values (0.05--0.82) for the bursts from the FRB\,20180301A source. The fact that these three repeat bursts were detected in the same observing session and exhibit unusual broad-band spectral extents also raises the possibility of a modification in the underlying emission mechanism. 

\section{Conclusions}\label{sec:conclusion}
The origin of overall FRB emission remains elusive. Repeating FRBs allows us to study the burst features in detail not possible with the apparent non-repeating ones. These repeating FRB sources exhibit a wide range of burst activity rates, reflecting the complexity and variability of these sources and much focus have been on the active repeating ones lately. Great diversity in the burst properties of repeating FRBs is emerging with dedicated long-term sensitive observations. The UWL instrument on the Parkes radio telescope provides a great opportunity to reveal wide-band nature of FRB emission in detail in regards to the signal spectrum and polarization. 

Over a monitoring period of two and a half years, we have detected $46$ bursts from the source of FRB\,20180301A. Our analysis shows significant Faraday RM variation with a sign reversal in the bursts. We also observed strong evidence of spectral depolarization at low frequencies (< 1.2\,GHz). Furthermore, we have discovered a secular evolution in the observed DM, with a variation rate of $-2.7\pm0.2\,{\rm pc\,cm^{-3}\,yr^{-1}}$. Notably, we detected repeat bursts with extremely narrow temporal widths of $\sim100\upmu$s and unusually wideband spectral envelopes of $\sim$1\,GHz, which suggests the presence of a sub-population of such bursts. The observed Faraday RM variation, spectral depolarization, and secular DM evolution provide important clues for understanding the FRB emission mechanisms and the local dynamic environments. However, more observations are needed to fully characterize the long-term variation trend and probe at lower and higher frequencies to explore the depolarization effects. A larger sample of bursts, similar in scale to FRB\,20121102A and FRB\,20201124A, would provide further insights into the time-variable burst activities and help identify any long-term trends necessary for characterizing the progenitor evolution. It could also allow for a search for underlying burst periodicity.

The growing league of well-studied repeating FRBs, including 20121102A, 20180916B, 20201124A, and 20190520B, has emerged as a powerful tool for exploring the physics of extreme environments. This study has successfully added FRB\,20180301A to the expanding list, further highlighting the commonalities among repeating FRBs and their dynamic progenitor environments. One such feature is the RM variation, which has been now observed in most repeating sources, showing both stochastic and secular regimes of change. The strong correlation of linear polarization with observing frequency, coupled with the significant time variability of CP and PA swings, suggests a complex and diverse emission geometry. These observations also demonstrate the need for more sensitive instruments with wide-band coverage to detect and reveal even more rich features of repeat bursts and their evolution over time.

\section*{Acknowledgements}
PK acknowledge support through the Australian Research Council (ARC) grant FL150100148.
RMS acknowledges support through ARC Future Fellowship FT190100155; ATD and RMS acknowledge support through ARC Discovery Project DP220102305.
SB is supported by a Dutch Research Council (NWO) Veni Fellowship (VI.Veni.212.058).
This work was performed on the OzSTAR national facility at Swinburne University of Technology. The OzSTAR program receives funding in part from the Astronomy National Collaborative Research Infrastructure Strategy (NCRIS) allocation provided by the Australian Government.
The Parkes radio telescope (\textit{Murriyang}) is part of the Australia Telescope National Facility (\url{https://ror.org/05qajvd42}) which is funded by the Australian Government for operation as a National Facility managed by CSIRO. We acknowledge the Wiradjuri people as the Traditional Owners of the Observatory site.
This research has made use of NASA's Astrophysics Data System Bibliographic Services, Astronomer's Telegram and software packages, including: \software{MATPLOTLIB} \citep{Hunter:2007_matplotlib}, \software{ASTROPY} \citep{Astropy:2013, Astropy:2018}, \software{NUMPY} \citep{Harris:2020_numpy}, \software{LMFIT} \citep{Newville:2016}, \software{PyMultiNest} \citep{Buchner:2014} and \software{CMASHER} for colormaps \citep{Velden:2020_cmasher}.

%%%%%%%%%%%%%%%%%%%%%%%%%%%%%%%%%%%%%%%%%%%%%%%%%%
\section*{Data Availability}
The data underlying this article will be shared on reasonable request to the corresponding author. Raw data from the Parkes telescope are archived on the CSIRO data access portal (\url{https://data.csiro.au}). 



\bibliographystyle{mnras}
\bibliography{references} % if your bibtex file is called example.bib

%%%%%%%%%%%%%%%%% APPENDICES %%%%%%%%%%%%%%%%%%%%%
%\appendix

%%%%%%%%%%%%%%%%%%%%%%%%%%%%%%%%%%%%%%%%%%%%%%%%%%

% Don't change these lines
\bsp	% typesetting comment
\label{lastpage}
\end{document}